\documentclass[a4paper,10pt, openany]{article}

\usepackage[cp1252]{inputenc}
\usepackage{amsthm}
\usepackage{amsmath}
\usepackage{amssymb}
\usepackage{amsfonts}
\newtheorem{definition}{Definition}
\newtheorem{theorem}{Theorem}
\newtheorem{proposition}{Proposition}

\newcommand{\B}[1]{\overline{#1}}
\newcommand{\ts}{\otimes}
\newcommand{\T}[1]{\widetilde{#1}}
\newcommand{\D}{\nabla_g \cdot}

\begin{document}

\title{A mathematicians' view of geometrical unification of classical physics in high-dimensional space-time. \\
\emph{L'Univers sans Foi ni Loi.}}
\author{Michel et Benoit Vaugon, Stephane Collion, Marie Dellinger, Zoe Faget.\\ \sl \small Institut de Math{\'e}matiques, Universit{\'e} Paris VI, Equipe
  G{\'e}om{\'e}trie et Dynamique, \\ \sl \small 175 rue Chevaleret, 75013
  Paris.  \\ \sl \small email: vaugon@math.jussieu.fr, stephane.collion@wanadoo.fr}
\date{August 2012}
\maketitle

\begin{abstract}

We propose in this paper a mathematicians' view of the Kaluza-Klein idea of a five dimensional space-time unifying gravitation and electromagnetism, and extension to higher-dimensional space-time. By considering the classification of positive Einstein curvature tensors and the classical Cauchy-Choquet-Bruhat theorems in general relativity, we introduce concepts of types and rigidity. Then, abandoning the usual requirement of a Ricci-flat five dimensional space-time, we show that a unified geometrical frame can be set for gravitation and electromagnetism, giving, by projection on the classical 4-dimensional space-time, the known Einstein-Maxwell-Lorentz equations for charged fluids. Thus, although not introducing, at least at this stage, new physics, we get a very aesthetic presentation of classical physics in the spirit of general relativity. The usual physical concepts, such as mass, energy, charge, trajectory, Maxwell-Lorentz law, are shown to be only various aspects of the geometry, for example curvature, of space-time considered as a Lorentzian manifold; that is no physical objects are introduced in space-time, no laws are given, everything is only geometry. We will then extend this setting to more than 5 dimensions, giving a precise mathematical frame for possible additional physical effects, preserving gravitation and electromagnetism. Version 22 04 2013.
\end{abstract}

\pagestyle{myheadings}\markboth{L'Univers sans Foi ni Loi}{L'Univers sans Foi ni Loi.}

\newpage
\tableofcontents
\newpage 

\section{Introduction}
We present in this paper a new vision on classical Kaluza-Klein theory, from a mathematician perspective.

We start our work by considering two aspect of classical general relativity, first the classification of possible energy-momentum tensors, that is, the classification of symmetric two-tensors satisfying certain positivity conditions, and second, the Cauchy-Choquet-Bruhat-type theorems. Building on these two concepts, we then define the notions of type and rigidity. It will lead to the important theorem 3 of section 4.2 showing the limits of the 4-dimensional setting. 

This will then lead us to a presentation of a possible setting for a five-dimensional space-time, in which the motion of a charged massive fluid appears as the projection on classical four dimensional space-time of a geodesic free fall. More precisely, defining in a purely geometrical manner a natural generalization of perfect fluid, we will show that the natural trajectories associated to it will exactly satisfy the Einstein-Maxwell-Lorentz equation of motion, as well as the classical equations of conservation of mass (baryonic number) and electrical charge. These equations will be simple and purely geometrical consequences of the Bianchi identities. Our main results in this direction will be summed up in theorem 7 of section 6.4.

We shall then see extensions of these models and ideas to more than 5 dimensions for space-time, introducing very nice (we think) models for higher dimensional fluids as well as 5+m-dimensional space-time containing Newtonian and/or electromagnetic potentials, giving new exact solutions to empty space equations, and showing possible extensions of Schwarschild solutions in a more general context. This will give a precise mathematical setting for describing possible additional physical effects, preserving gravitation and electromagnetism. See section 7. The climax will be the results of section 7.3, theorems 9 and 10, and the examples of section 7.4. Note that in section 7.2 we shall give an alternative definition of a fiber bundle, which insists on the fact that it is really a structure imposed on the total space and not on the base space, giving in particular a natural and intrinsic notion of "small dimensions" as fibers. This will be more appropriate for the physical use we have for this structure, as we consider the total space to be the "real" universe, and the base space only an approximate model, limited to restricted possible physical measurements. 

Sections 2 and 3 are a translation of the last chapter of a General Relativity course written by Michel Vaugon. They will however introduce the ideas on which the paper is based.

Our main notations and conventions can be found at the end of the paper.
\newpage

\section{Motivations}
In the beginning was spacetime and freedom.\\
Then came the physicists, their laws and their machines. \\

The axiomatism of General relativity is beautiful and very simple, so long as one does not attempt to fully introduce electromagnetism in it.

Indeed, with the help of sophisticated mathematical tools we can cross the
bridge from classical mechanics to Special Relativity then to General
Relativity while reducing the number of physical principles and laws at the
same time. For example, ambiguous concepts such as absolute time, Galilean
observers, or straight line disappear. 
General Relativity is the natural conclusion of this process, giving the
Universe the most general frame, free of {\it ad hoc} concepts and useless
principles.   
Only the Lorentzian metric survives. But from a mathematical point of view,
differential and Riemmannian geometry are globally captured, in particular 
Bianchi's second identity which is a key result despite its
mathematical simplicity.

One of the most basic physical concept that we absolutely need to get rid of
when working in General Relativity is the idea that one introduces matter in
space-time, seen as the framework of physical experiments. Matter and
geometry co-exist, as it is expressed in Einstein's equation. However, even
though it was Einstein's fundamental idea, centuries of physical experiments
consisting in introducing objects in an experimental framework (marbles on a rampe,
electrons in a magnetic field\ldots) failed to bring this idea to its
final outcome. 	Indeed, to this day,
most General Relativity courses introduce matter in the space-time Lorentzian manifold modelized by the energy-impulsion
tensor, the link between matter and curvature being given by Einstein's
equation.
    
 One remarkable fact (that actually  inspired
 Einstein's equation) is that the  Bianchi's second identity applied to the left
 part of this equation (representing Einstein's curvature) coincide with the
 law of mass conservation applied to the right member of the equation,
 representing matter (a fluid).
 
This coincidence suggests another vision on the axiomatic: matter's physical data can
be characterized by the geometry. Precisely, physical objects that humans want to
describe are just expressions of the manifold's geometry, in particular the
space-time curvature. Even more precisely, data such as density of energy,
density of mass, pressure, are defined as characteristics of the curvature and
it is then Bianchi's identity alone that gives Einstein's equation.

For the moment, we can summarize General Relativity's axiomatic 
as follow:
     
A. Space-time is a Lorentzian manifold of 4 dimensions. (Observers,
propre time, and space seen by an observer are defined as usual). 
      
B. We canonically define data based on the Lorentzian manifold's curvature
tensor which will physically represent : density of energy, density of mass of
a fluid, pressure of a fluid, unit vector of fluid curves, etc\ldots

No physical object is added : there is only geometry. 

No law is added. Bianchi's second identity gives mass conservation law (when
appliable), the fact that for a perfect dust fluid, curves are geodesic, the
equation verified by a perfect fluid, etc\ldots

Hence this identity gives us : an approximation of classical mechanic
(gravitation), big bang and big crunch for an isotrop and homogenous domain,
the study of spherical symmetry in space (Schwarschild) and therefore movements
of planets, light deviation, black holes.
We precisely find all general relativity applied to perfect fluids.

Unfortunately, this vision can not deal with electromagnetism. Indeed, even
though we can define in a canonical manner the energy-impulsion tensor
representing electromagnetism, we can not find a canonical definition for the
2-form of electromagnetism and Maxwell-Lorentz equations.
It is this precise problem that precursors Einstein, Weyl, Kaluza, Klein,
Rainich have attempted to solve. We'll get back to this point later on.
      
To summarize, the study of electrically neutral perfect fluid in general relativity can be
reduced to the study of Lorentzian manifolds. In other words, physical laws
regarding matter fluids are just translations of Riemannian geometry theorems.
However, it is not the case for electromagnetism which needs  the introduction
of an exact 2-form verifying  ``laws'', namely
Maxwell equations, in the space-time manifold, in order to have a
formal definition in general relativity.

To conclude this introduction, it is interesting to note that there is no
``general principle'' in the general relativity's axiomatic, contrary to
classical physics or special relativity. 
Indeed, these two domains need the assumption of an homogeneous and isotrop
space-time, as well as law invariance with respect to Galilean observers. This
translates mathematicaly to laws invariance under the action of a Galilean or a
Lorentz group. None of this remains in general relativity. Homogeneity,
isotropy, or more generally invariance are just approximations allowing
approximate computations and are not general principles.
Ironically, one could note that those assumed principles lead to the formal
definition of general relativity's axiomatic, but they do not survive in the
definition.
\newpage

\section{Classical General Relativity.}

This section is chapter 15 of Michel V's lecture book. 

\subsection{Physical concepts based on the curvature tensor}

Physical objects such as energy's density, mass density, pressure, fluid's
curves, electro-magnetic fields, etc. were chosen to describe matter's behavior
following the evolution of physic's theories. They correspond to ``human's''
measurements (in the sense that they can be observed by humans) and are pertinent in
classical physics as well as special relativity, but are not necessarily well
adapted to general relativity. The following definitions may seem a
bit artificial (despite their canonical definitions) since they are guided by
the will to regain classical physic notions.

In different domains of space-time, we might find different ``physical objects''.
Indeed, we can easily believe that in certain domains there is no matter fluid
but only electromagnetic fields, or the opposite, or nothing at all.
The definition of physical data on a space-time domain will depend on the
Lorentzian structure of this domain, especially  Einstein's curvature
$$ G = \mathit{Ricc}- \frac{1}{2} R g. $$ 

We begin by a classification of domains with respect to their type of Einstein's
curvature. We limit ourselves to domains for which physical objects defined
from the curvature have already been observed. Furthermore, we will limit
ourselves, for simplificity, to domains in which only perfect fluids and
electromagnetism fields exist. Of course, if we pick a domain of a Lorentzian
manifold at random we might find a part of space-time that has never been
observed and it is then hard to guess which physical objects would be
interesting to define.
We do not ask ourselves this kind of questions here.
The selected domains are those satisfying the ``dominant energy condition''
because only those kind of domains have been studied so far.
\\

\textbf{Dominant energy condition} 
A Lorentzian manifold $(\Omega, g)$ of Einstein's curvature $G$ verify the
dominant energy condition if, for all timelike vectors $\vec{v}$, 
 $G(\vec{v},\vec{v}) \geq 0$
and if the energy-impulsion density vector defined by $^eG(\vec{v}) $
is spacelike or isotropic, where $^eG$ is the endomorphism associated to $G$ by $g$.
\\

Physically, this condition means that for all observers 
 $\vec{v},$ the density of energy seen by $\vec{v}$
 is always positive and the energy-impulsion density vector seen by this
 observer does not exceed the speed of light.
We roughly translate this last point by saying that no information associated to an 
energy flow can exceed the speed of light in $\Omega$.
 Last, let us note that in traditional general relativity literature the
 dominant energy condition is presented as an axiom, but we find no need
 for that and rather choose to believe it is possible for some space-time
 domains not to verify this condition.

\subsection{A classification of Lorentzian manifolds domains with respect to
their Einstein tensor}

A complete classification is given in Hawkings-Ellis \cite{HE},
but the presentation differs because we aim at using the classification in
order to {\bf define} traditional physical objects.

Let us note for a start that if $\vec{u_x}$ is a spacelike unit vector on
a Lorentzian manifold $(M,g),$ then  all tensors  $T $ twice covariant,
symmetric in $x$, can be decomposed \textbf{ in a unique way} in the following
fashion
\begin{equation}\label{1}
T_{ij} = A u_i u_j + B(g_{ij} + u_i u_j) + \Pi_{ij} + (q_i u_j+ q_j u_i) 
\end{equation}
where
\begin{itemize}
 \item  $A$ and $B$ are two real numbers
\item  $\vec{u_{x}} \in Ker \, ^e \Pi_x,$ where $ ^e\Pi$ is the endomorphism
 associated by $g$ to $\Pi$.
Hence we have $\Pi_{ij} u^i = 0$
\item $tr \, ^e\Pi_x = 0$ i.e $\Pi_{ij}g^{ij} =0$ 
\item $\vec{q_x}$ is $g$-orthogonal to $\vec{u_x}$, i.e $q_i u^i =
0.$\\
\end{itemize}

In particular, we get 
$$
A = T_{ij} u^i u^j \qquad B = \frac{1}{3} T_{ij} (g^{ij} + u^i u^j) \qquad
\mbox{ and }\qquad q^i = -T_{jk} u^j (g^{jk} + u^k u^i) $$

Of course, this decomposition of  $T$ is linked to the choice of $\vec{u_x},$ 
but as soon as
 $\vec{u_x}$ is canonically defined, $A_x, B_x,
 \Pi_x$ and $q_x$ are defined unambiguously. Furthermore in the case where
 $T$ is Einstein's curvature, they become the physical data that we are
looking for. 
 Domains in which $\vec{u_x}$ is canonical will be those for which a fluid
 exists ($\vec{u_x}$ will then be seen as the unit tangent vector to the fluid's
 curve). They match types 1 and 2 that we detailed next. In all cases, the
 common terminology for data defined by (\ref{1}) when $T = G$ is the
 following:

\begin{itemize}
 \item $A_x$ : energy density seen at point $x$ by observer $\vec{u_x}.$
 \item $B_x$ : pressure seen at point $x$ by observer $\vec{u_x}.$
 \item $\Pi_x$ : anisotropic pressure tensor seen at point $x$ by observer $\vec{u_x}.$
 \item $q_x$ : energy flow seen by $\vec{u_x}.$
\end{itemize}

As we said before, we limit ourselves to domains which contain (at most) only
fluids and electromagnetism. Einstein curvature's tensor is then one of the
following types :

\subsubsection*{Type 0 : Emptiness of matter and electromagnetism}
Domains where for all $x$ in the domain  $G_x = 0$  (equivalent to a
Ricci curvature equal to zero) represent parts of the space-time where there
is no fluid nor electromagnetism, nor anything else.
Beware that those domains may be geometrically complex and are especially
interesting : domains with spherical symmetry fall in this category, as well as
Schwarzshild model and non-charged black holes.

\subsubsection*{Type 1 : Non-charged fluids}
A domain is of type 1 if for all $x$ in the domain, $^eG_x$ has the following
properties:
\begin{itemize}
 \item $^eG_x$ has an eigenvalue $-\mu <0$ with eigenspace
 $\mathcal E_{-\mu}$, $dim(\mathcal E_{-\mu})=1$ and timelike.
\item $^eG_x$ has an eigenvalue $\lambda$ verifying $-\mu <
\lambda<\mu$, with eigenspace $\mathcal E_{\lambda}$, $dim(\mathcal
E_{\lambda})=3$, such that $\mathcal E_\lambda \perp_g \mathcal
E_{-\mu}.$
\end{itemize}
This is equivalent to the existence of a $g$-orthonormal base  satisfying
$$ \left( G^i_{\  j} \right) = \begin{pmatrix}
-\mu & 0&0&0\\
0 &  \lambda&0&0\\
0&0&\lambda&0\\
0&0&0&\lambda
\end{pmatrix}
\qquad \mbox{ with  } \mu >0 \mbox{ and } -\mu < \lambda < \mu.
$$
Such domains physically represent domains in which there is a fluid and nothing
else. We then {\bf define} unambiguously:
\begin{itemize}
 \item \textbf{The unit tangent vector $\vec{ u_x } $ to the fluid's
 curve} by the only unit vector in the orientation of $\mathcal E_{-\mu}.$
\item \textbf{The fluid's energy's density in $x$} by the real positive number
$\mu.$
 \item \textbf{The fluid's pressure in $x$ } by the real number $\lambda.$
\end{itemize}

In this case, considering (\ref{1}), $G_x$ can be split in a unique manner 
the following way:
$$G_{ij} = \mu u_i u_j + \lambda (g_{ij} + u_i u_j)$$

\textbf{Remark :} If $\lambda = 0,$  there is no pressure and we get a domain
where there is a {\bf truly perfect} fluid. In this case and by definition, the
energy's density is also the \textbf{ mass density }.

\subsubsection*{Type 2 : Perfect fluid and electromagnetic field}
A domain is of type 2 if for all $x$ of the domain, $^eG_x$ has the following
properties:
\begin{itemize}
 \item  $^eG_x$ has an eigenvalue $-\mu <0$ of eigenspace $\mathcal E_{-\mu}$,
 $dim(\mathcal E_{-\mu})=1$ and spacelike.
\item  $^eG_x$ has an eigenvalue $\lambda_1$ of eigenspace $\mathcal
E_{\lambda_1}$, $dim(\mathcal E_{\lambda_1})=1$, such that
$\mathcal E_{\lambda_1} \perp_g \mathcal E_{-\mu}.$
\item  $^eG_x$ has an eigenvalue $\lambda_2$ of eigenspace $\mathcal
E_{\lambda_2}$, $dim(\mathcal E_{\lambda_2})=2,$ such that
$\mathcal E_{\lambda_2} \perp_g (\mathcal E_{\lambda_1} \oplus \mathcal
E_{-\mu})$  and such that $-\mu < \lambda_1 <\lambda_2 <\mu.$
\end{itemize}
This is equivalent to the existence of a $g$-orthonormal base satisfying :
$$
\left( G^i_{\  j} \right) = \begin{pmatrix}
-\mu & 0&0&0\\
0 &  \lambda_1&0&0\\
0&0&\lambda_2&0\\
0&0&0&\lambda_2
\end{pmatrix}
\qquad \mbox{ with  } \mu >0 \mbox{ and } -\mu < \lambda_1 < \lambda_2 < \mu.
$$
Such domains physically represent the association of a perfect fluid and an
electromagnetic field.\\
We then \textbf{define} unambiguously:
\begin{itemize}
 \item \textbf{The unit tangent vector $\vec{ u_x }$ to the fluid's
 curve} by the only unit vector in the orientation of $\mathcal E_{-\mu}.$
\item \textbf{The fluid's energy density in $x$ } by the positive real number
$\mu$ (sum of the fluid's energy density and the electromagnetic's
energy's density defined below).
\item \textbf{The fluid's energy density in $x$} by the positive real
number $\mu - \frac{1}{2}(\lambda_2 - \lambda_1).$
\item \textbf{The electromagnetic energy density in $x$} by the positive
real number $ \frac{1}{2}(\lambda_2 - \lambda_1).$
 \item \textbf{The fluid's pressure in $x$} by the real number $\frac{1}{2}(\lambda_1+\lambda_2).$
 \item \textbf{The electromagnetic pressure in $x$} by the real number
 $\frac{1}{6}(\lambda_2 - \lambda_1).$
\item \textbf{The electromagnetic tensor in $x$} by $\Pi_{ij}$ with trace
equal to zero, given by (\ref{1}) when $T=G.$
\end{itemize}
Justification of those choices will be given in section \ref{justif}.

In this case, considering (\ref{1}), $G_x$ can be decomposed in a unique manner
the following way: 
$$ G_{ij} = \mu u_i u_j + \left(\frac{\lambda_1+\lambda_2}{2} + \frac{\lambda_2 - \lambda_1}{6}   \right) (g_{ij} + u_i u_j) + \Pi_{ij}
$$

{\bf Remarks :} In the equality above, we have $\Pi_{ij} g^{ij} = 0.$ We also
note for this type 2 that there exists  a spacelike vector $\vec{v_x}$,
canonically determined up to orientation by $\mathcal E_{\lambda_1}.$

For the following types, we no longer deal with fluids, only with an
electromagnetic field.
\subsubsection*{Type 3 : An example of an electromagnetic field in an empty space}
A domain is of type 3 if for all $x$ in the domain, $^eG_x$ has the following
properties: 
\begin{itemize}
 \item $^eG_x$ has an eigenvalue $-\mu <0$ with an eigenspace
 $\mathcal E_{-\mu}$, $dim(\mathcal E_{-\mu})=2$.
\item $^eG_x$ has an eigenvalue $\lambda = \mu$ with a spacelike eigenspace
$\mathcal E_{\lambda}$,  $dim(\mathcal E_{\lambda})=2$, such that $\mathcal
E_{\lambda} \perp_g \mathcal E_{-\mu}.$
\end{itemize}
This is equivalent to the existence of a $g$-orthonormal base verifying:
 $$
\left( G^i_{\  j} \right) = \begin{pmatrix}
-\mu & 0&0&0\\
0 & -\mu&0&0\\
0&0&\mu&0\\
0&0&0&\mu
\end{pmatrix}
\qquad \mbox{ with } \mu >0.
$$
Such domains physically represent an example of an electromagnetic field in a
space empty of matter, i.e in the absence of fluids.
The Reissner-Nordstr\"om model [H-E] gives an example in spherical symmetry
 (static electromagnetism). There is no canonical split of $G_x$ in this case.
 Indeed, such a split is linked to the choice of a spacelike vector $\vec{u_x}$,
 and in this type it is not possible to define such a vector canonically. Canonical datas of this type are the real number $\mu$ and the two eigenspaces
$\mathcal E_{-\mu}$ and $\mathcal E_{\mu}.$

\subsubsection*{Type 4 : Electromagnetic wave}
A domain is of type 4 if for all $x$ in the domain, $^eG_x$ has the following
properties:
\begin{itemize}
 \item $^eG_x$ has a single eigenvalue $0$ of eigenspace $\mathcal E_{0},$
 $dim(\mathcal E_{0})=3$, $\mathcal E_0$ does not contain any timelike
 vector, isotrop vectors of $\mathcal E_0$ form a sub-space of
 dimension 1 ($\mathcal E_0$ is actually tangent to the light cone).
\end{itemize}
This is equivalent to the existence of a $g$-orthonormal base verifying:
$$
\left( G^i_{\  j} \right) = \begin{pmatrix}
-1 & 1&0&0\\
-1 & 1&0&0\\
0&0&0&0\\
0&0&0&0
\end{pmatrix}
$$

{\bf Remark :} The choice of the first element (spacelike vector) in the
$g$-orthonormal base is not canonical. Indeed, it can be chosen in a way such
that $$\left( G^i_{\  j}
\right) = \begin{pmatrix} -\nu & \nu&0&0\\ -\nu & \nu&0&0\\
0&0&0&0\\
0&0&0&0
\end{pmatrix}$$ where $\nu >0.$

Such domains physically represent domains of space-time where one can find only
electromagnetic waves and no fluids. We then define unambiguously an isotropic
oriented line (representing the propagation direction of the electromagnetic
wave), given by the sub-space of dimension 1 formed by the isotropic
vectors of $\mathcal E_0.$ If one choses a non-zero, in orientation, isotrop
vector $\vec{i}\in \mathcal E_0,$ then $G$ seen as twice contravariant can be
written : $G = \nu \vec{i} \otimes \vec{i}$ with $\nu>0.$ 
This writing is not like (\ref{1}), and of course $\nu$ relies on the
choice of $\vec{ i}$ (itself defined up to a product by a real number).

{\bf Remark:} The case $G_x = -\mu Id$ with $\mu>0$ is a limit case of type
1. It is compatible with the dominant energy condition and matches the case of
incompressible fluids, see Choquet-Bruat \cite{CB}

\subsection{In $4$ dimensions, geometry is not sufficient to describe
electromagnetism.}

\subsubsection{Justification for choices of types 2, 3 and 4 \label{justif}}

We chose to limit ourselves to cases containing electromagnetism, and perhaps
a fluid as in type 2. For this, we started with the classical presentation of
electromagnetism in general relativity, which introduces in the Lorentzian
manifold an exact 2-form $F$, and we assumed that the
corresponding energy-impulsion tensor (which adds itself to the fluid's
energy-impulsion tensor for type 2) can be written as 
$$
T_{ij} = F_{ik} F^k_{\ j} + \frac{1}{4} F_{kl} F^{kl} g_{ij}.
$$
We further assume when a fluid is present that vectors tangent to the
fluid's lines are eigenvectors of $^eG$ (other cases are much longer to
describe).
Hence, considering the special form of the tensor $T_{ij},$ and assuming the
dominant energy condition, quick calculations show that only types 2, 3 and 4
are possible. Of course, with our new look on the axiomatic presented in this
section, types 2, 3, 4 are the ones allowing us to define all data of
electromagnetism (chosen specifically so that we recover classical
notions).

\subsubsection{Electromagnetism is not purely geometrical in 4 dimension.}
However, domains of types 2, 3 or  4  allow us to define
\textbf{only} the following tensors: 
\begin{eqnarray*}
T_{ij} &=& \frac1{2} (\lambda_2 - \lambda_1) u_i u_j + \frac{1}{6} (\lambda_2 -
\lambda_1) (g_{ij} + u_i u_j) + \Pi_{ij} \quad \mbox{  for type } 2 \\ &&\\
T_{ij} &= &G_{ij} \quad \mbox{  for types } 3 \mbox{ and } 4. 
\end{eqnarray*}
For each of those cases, tensors correspond to the classical
energy-impulsion tensor of electromagnetism, but they \textbf{don't  allow to
retrieve the electromagnetism 2-form $F$ canonically} (and a fortiori
Maxwell's equations). One can note that for a given symmetrical tensor $T_{ij}$, there
exists in general an infinite number of anti-symmetrical tensors $F_{ij}$ such
that $ T_{ij} = F_{ik} F^k_{\ j} + \frac{1}{4} F_{kl} F^{kl} g_{ij}.$
\textbf{Therefore, it is not possible to retrieve classical electromagnetism,
({\it i.e} the $2$-form $F$ and Maxwell equations) with only 
the Lorentzian manifold's geometry as a given.} However, we can wonder if the
energy-impulsion tensor $T_{ij}$ of electromagnetism is sufficient to describe
 physical reality, in particular a fluid's behavior
 (since, in the end, only fluids are physically observable).
 The answer is still no. The opposite would mean we could describe electromagnetic phenomenons without having to use the  $2$-form $F,$ in other worlds
without using the electromagnetic field. We will show precisely in section 2.7
that the knowledge of the tensor $T_{ij} $ alone can not lead to a physical
theory sufficiently deterministic, contrary to the classical theory of
electromagnetism in general relativity (which consists in introducing the
2-form $F$ with its energy-impulsion tensor and Maxwell equations in the
space-time).
 
\noindent \textbf{Therefore, in 4
dimensions, one can not describe electromagnetism using the
Lorentzian manifold's geometry alone.}

\subsection{Where we verify that fundamental laws on fluids can be deduced
from Bianchi's identity}
We consider a domain of space-time of type $1$, {\it i.e} containing only a
non-charged fluid. We have $$
G_{ij} = \mu u_i u_j + \lambda (g_{ij} + u_i u_j) \qquad \mbox{ with } \mu>0
\mbox{ and } -\mu < \lambda < \mu. $$

\subsubsection{Case of a truly perfect fluid:
$\lambda = 0$} In this case,  $\mu$ is the mass density. Bianchi's identity
gives us $$ 0 = \nabla^i G_{ij} = \nabla^i
\left(\mu u_i u_j \right)  = \nabla^i \left(\mu u_i \right) u_j + \mu u_i \nabla^i \left( u_j \right) $$
Since $u^j u_j = -1,$ we have $0= \nabla^i(u^j) u_j  + u^j \nabla^i(u_j) = 2 
u^j \nabla^i(u_j)$ hence $ u^j \nabla^i(u_j) =0.$ Therefore $$
0 = u^j \nabla^i(G_{ij}) = - \nabla^i(\mu u_i).
$$ 
In other words, the divergence of the vector field $\mu \vec{u}$ equals zero.
This, along with Stokes theorem, gives the mass conservation law.\\
We also deduce $0= u_i \nabla^i u_j, $ which means $D_{\vec{u}} \vec{u}=0.$
This means exactly that fluids curves, parametrized such that  $\vec{u}$ is
the tangent vector everywhere, are geodesics.

\subsubsection{The more general case of an ``isentropic" perfect fluid}
The definition of a fluid given by type $1$ domains does not allow to
introduce the notion of mass density based on the notions of
density of energy and pressure. This can be physically interpreted by saying
that in those fluids there is a notion of ``internal energy'' depending on the
nature of the fluid. In fact, energy can transform into mass and vice-versa
throughout the fluid's evolution, and a mass conservation law can be
invalidated in the general case. The only ``conservation law'' still valid is
Bianchi's identity : $\nabla^i G_{ij} = 0.$ 
The notion of mass density can actually be defined for certain categories
of perfect fluids, for example such as isentropic perfect fluids (truly perfect
fluids are a special case). An isentropic perfect fluid is a perfect fluid for
which there exists a differentiable function $\epsilon : \mathbb{R} \rightarrow \mathbb{R}$
(called elastic potential function) depending on the nature of the fluid,
for which the two following statements (inspired by classical fluid mechanics)
hold :
\begin{itemize}
 \item $(i)$ $\mu = \rho (1 + \epsilon(\rho))$
\item $(ii)$ $\lambda = \rho^2 \epsilon'(\rho)$
\end{itemize}
where $\mu $ is the density of energy, $\lambda$ is the pressure and $\rho$ is
the fluid's mass density. Since the mapping $x \mapsto x(1+\epsilon(x))$ is
assumed to be reversible, $(i)$ fully determines 
$\rho $ from $\mu$ and $\epsilon.$ Using $(i)$ et $(ii)$, $\epsilon $
also gives an equation linking $\lambda$ and $\mu,$ called the state equation
of the perfect fluid (also depending on the fluid's nature because of
$\epsilon$).

Hence from $(i)$ and $(ii)$ we get 
$$\frac{d\mu}{d\rho} = 1+\epsilon(\rho)+\rho \epsilon'(\rho) \qquad \mbox{ then
} \qquad \rho \frac{d\mu}{d\rho} = \lambda + \mu$$  Bianchi's identity gives
\begin{equation}\label{2}
\begin{split}
0 = \nabla^i G_{ij} &= \nabla^i \left( \mu u_i u_j+ \lambda (g_{ij} + u_i u_j) \right)\\
&= \nabla^i(\lambda+\mu) u_i u_j + (\lambda+\mu) \nabla^i(u_i u_j) + (\nabla^i \lambda) g_{ij}
\end{split}
\end{equation}
If we do a contracting product with $u^j,$ and knowing that $u_j
u^j = -1$ and $u^j \nabla^i u_j = 0,$ we get :
\begin{equation}\label{3}
(\nabla^i \mu) u_i+ (\mu + \lambda) \nabla^i u_i = 0
\end{equation}
But $\nabla^i \mu = \frac{d\mu}{d\rho } \nabla^i \rho,$ hence $\rho \nabla^i \mu
= (\lambda+ \mu) \nabla^i \rho.$ From that, we get $$(\lambda+\mu) (\nabla^i
\rho) u_i + (\lambda + \mu)\rho \nabla^i u_i = 0 $$ and finally, since
$\lambda+\mu>0 : $ 
$$
\boxed{ \nabla^i (\rho u_i) =0}
$$
which is once again, the mass conservation law.\\

{\bf Remark :} In the case of a perfect fluid, the fluid's curves are not
necessarily images of geodesics.
\\

Finally, getting back to (\ref{2}) and (\ref{3}), we get the ``perfect fluid's
equation" : 
$$
(\mu + \lambda ) u_i \nabla^i u_j + (g_{ij} + u_i u_j) \nabla^i \lambda=0
$$ 

\newpage

\section{Causality in general relativity. Rigidity in Lorentzian geometry.}

 \subsection{Causality in general relativity.}
 Let us consider an observer in general relativity. We suppose that he was able to record all physical data on an open subset of space-time. If this open subset contains a Cauchy hypersurface and if Cauchy-type theorems apply to these data, then the observer can deduce all the physical data in the domain of dependence of this Cauchy hypersurface (which contains the open subset under consideration). But in fact, physically, the observer cannot have knowledge of the physical data on this hypersurface unless he is at the "future temporal limit" of the domain of dependence; so he in fact has to assume the data on a larger Cauchy hypersurface. In fact only data on the hypersurface are required, not on the all open subset. However data on a Cauchy hypersurface required to apply Caychy theorems are difficult to define. For our purpose, it will be easier to assume that "initial conditions" are in fact data on an open subset of $(M,g)$, (which in fact is more close to a "physical" reality.) We are not here concerned with the minimality of these conditions. 
 
 If we consider that physical data are only geometric characteristics of the Lorentzian manifold (of its curvature), the problem seems to be the following. Under what conditions does the knowledge of the metric tensor $g$ on an open subset $\Omega$ of $M$ completely determines $g$ on the domain of dependence of $\Omega$ ? This question, thus posed, is not precise, as $\Omega$ is already supposed to be an open subset of a Lorentzian manifold $(M,g)$ where the tensor $g$ is already known, as this is needed to even define the domain of dependence. A more precise question would be the following: Let $(\Omega ,g)$ be a Lorentzian manifold. For every inextendible Lorentzian manifold $(M,\tilde g)$ (i.e. possible space-times) in which $(\Omega, g)$ can be isometrically embedded, are the domains of dependence of $\Omega$ all isometrical ? In fact to set the problem, it is not really necessary to assume the manifolds to be inextendible. The next section will expose precisely the definitions and the results obtained concerning these problems. The answers will be presented as \emph{rigidity theorems}. These are purely mathematical theorems concerning Lorentzian manifolds, (they have no meaning for Riemannian manifolds, as we need the concept of domain of dependence, which is purely Lorentzian).
 
 \subsection{Rigidity theorems for Lorentzian manifolds.}
 
The following definitions are valid on Lorentzian manifolds of dimension $n\geqslant4$.

\begin{definition}
Let $(\Omega, g)$ be a Lorentzian manifold, or a domain of such a manifold. To define a \emph{type} on $(\Omega, g)$ is to define properties $P(g)$ that depend only on the geometry of $(\Omega, g)$, that is properties that are left invariant by any isometry of $\Omega$ : $P(\varphi^{*}g)=P(g)$ for any isometry $\varphi$ on $\Omega$. It is important to note that an isometry being also an homeomorphism and a diffeomorphism, $P(g)$ also depends on the topological and the differential structure of $\Omega$.
\end{definition}

For manifolds of dimension 4, we can consider the types described in the previous section, and to separate domains of space-time where there is electromagnetism from those where there isn't, we shall say that a domain $\Omega$ is: \\
-of type I if $\forall x \in \Omega$, $G_{x}$ is of type 0 or 1,\\
-of type II if $\forall x \in \Omega$, $G_{x}$ is of type 2, 3 or 4.

We shall write I or II or 1,2,... instead of P(g) in these cases. To make it short, we shall write X for a type of any sort.

\begin{definition} A \emph{causal} domain $(\Omega, S, g)$ is given by a time-oriented manifold $(\Omega, g)$ with a space like hypersurface $S$ that is a Cauchy hypersurface for $(\Omega,g)$; $S$ is supposed to be of class $C^{2}$. A causal domain of type $P(g)$ is a causal domain $(\Omega, S, g)$ where $(\Omega, g)$ is of type $P(g)$.  We will denote this by writing $(\Omega, g)$ is a $DC(P(g))$.
\end{definition}

Remark: in this subsection, an isometry from a Lorentzian manifold $(\Omega_{1}, g_{1})$ to a Lorentzian manifold $(\Omega_{2}, g_{2})$ is only defined to be a smooth function $\varphi : \Omega_{1} \rightarrow \Omega_{2}$ such that $\varphi^{*}(g_{2})=g_{1}$ (we can also suppose that it preserves time orientation).

\begin{definition} 
Let $(\Omega_{1},S_{1})$ and $(\Omega_{2},S_{2})$ be two $DC(X)$. An isometry $\Theta$ from $(\Omega_{1},S_{1})$ into $(\Omega_{2},S_{2})$ is a surjective isometry from $\Omega_{1}$ into $\Omega_{2}$ such that $\Theta(S_{1})=S_{2}$.
\end{definition}

\begin{definition}
A X-extension of a $DC(X)$ $(\Omega, S)$ is a couple $(\Theta_{1}, \Omega_{1})$ where $\Theta_{1}$ is an injective isometry (not necessarilly surjective) from $\Omega$ into $\Omega_{1}$ such that $(\Theta_{1}(S),\Omega_{1})$ is a $DC(X)$.
\end{definition}

\begin{definition}
Let $(\Theta_{1}, \Omega_{1})$ be a X-extension of $(\Omega, S)$. A $X$-overextension of $(\Theta_{1}, \Omega_{1})$ is a $X$-extension $(\Theta_{2}, \Omega_{2})$ of $(S,\Omega)$ such that there exists an injective isometry $\psi$ from $\Omega_{1}$ into $\Omega_{2}$ that satisfies on $S$ : $\psi \circ \Theta_{1} = \Theta_{2}$.  Then, thanks to the following proposition, $\psi \circ \Theta_{1} = \Theta_{2}$ on $\Omega$.
\end{definition}

\begin{proposition}
Let $\Omega$ and $\Omega_{1}$ be two Lorentzian manifolds, $S$ a Cauchy hypersurface for $\Omega$, $\varphi_{1}$ and $\varphi_{2}$ two injective isometries from $\Omega$ into $\Omega_{1}$. Then, if $(\varphi_{1})_{ |S} = (\varphi_{2})_{ |S}$, we have $\varphi_{1} = \varphi_{2}$.
\end{proposition}

\emph{Proof : } Let $x\in \Omega$. There exists a unique $x_S \in S$ such that the unique geodesic curve, timelike or light like, joigning $x$ to $x_S$ is $g\bot$ to $S$ at the point $x_S$, (it is, for timelike curves, the curve with the maximum proper time between $x$ and $S$, see [H-E]). Let $C$ be this geodesic (well parametred), then, as $\phi_1$ and $\phi_2$ are isometric embeddings, $\phi_1 \circ C$ and $\phi_1 \circ C$ are geodesics orthogonal to $\phi_1(S)$, resp. to $\phi_2(S)$, at point $\phi_1(x_S)$, resp. $\phi_2(x_S)$. Because $x_S\in S$, $\phi_1(x_S)=\phi_2(x_S)$ and because there exists a unique geodesic curve orthogonal to $\phi_1(S)=\phi_2(S)$ at point $\phi_1(x_S)$, we have $\phi_1 \circ C=\phi_1 \circ C$ on $[0,t_0]$, therefore $\phi_1 \circ C(t_0)=\phi_1 \circ C(t_0)$, that is $\phi_1(x)=\phi_2(x)$.

\begin{definition}
An isometry $\varphi$ from a $X$-extension $(\Theta_{1}, \Omega_{1})$ of $(\Omega, S)$ into a $X$-extension $(\Theta_{2}, \Omega_{2})$ of $(\Omega, S)$ a bijective isometry from $\Omega_{1}$ into $\Omega_{2}$ such that : on S, $\varphi \circ \Theta_{1} = \Theta_{2}$. (Then $\varphi \circ \Theta_{1} = \Theta_{2}$ on $\Omega$).
\end{definition}

\begin{definition}
Let $(\Theta_{1}, \Omega_{1})$ be a $X$-extension of $(\Omega, S)$. $(\Theta_{1}, \Omega_{1})$ is a maximal $X$-extension of $(S, \Omega)$ if all $X$-overextension of $(\Theta_{1}, \Omega_{1})$ is isometric to $(\Theta_{1}, \Omega_{1})$.
\end{definition}

\begin{theorem}
Let $(S, \Omega)$ be a $DC(X)$. There always exists a maximal $X$-extension of $(S, \Omega)$. (There can exist more than one, not isometric one to another.)
\end{theorem}

\emph{Proof}: Difficult. One need to use Zorn axiom. See Hawking-Ellis.

\begin{theorem}
(Rigidity for the $DC(X)$ when X=0,1 or I). Let $(S, \Omega)$ be a causal domain of type $X=$0,1 or I. The maximal X-extensions of $(S, \Omega)$ are all isometric one to another.
\end{theorem}

\emph{Proof}. One applies the Cauchy theorems of general relativity with initial conditions on S translated by using the fact that the domain is of type 0,1 or I. (See Choquet-Bruhat, with some unicity theorems to be proven).

Applying Choquet-Bruhat theorems with appropriate initial data on $S$, we can state the following theorem, although complete technical details have yet to be written down.

\begin{theorem}
(The DC(X) are not rigids when X=2,3,4 or II.) Let $(S, \Omega)$ be a causal domain of type X=2,3,4 or II such that $(Id, \Omega)$ is not X-maximal (where $Id$ is the identity on $\Omega$.) There exist an infinity of $X$-maximal extensions of $(S, \Omega)$ not isometric one to another.
\end{theorem}

\emph{Remark : }If type X $\Rightarrow$ type Y, (for example, as one can see from the definitions, type 0 $\Rightarrow$ type 1, type 1 $\Rightarrow$ type I, type 2 $\Rightarrow$ type II, etc...), then, a X-extension $(\Theta_{1}, \Omega_{1})$ of $(\Omega, S)$ is also a Y-extension of $(S, \Omega)$. One can therefore, for a given X-extension $(\Theta_{1}, \Omega_{1})$ of $(\Omega, S)$, consider X-overextensions and Y-overextensions of $(\Theta_{1}, \Omega_{1})$, which have no reasons of being the same (up to isometry). We however have the following theorem (Hawking-Ellis, lemma 4.3.1), translated here in our language (whose proof is not easy).

\begin{theorem}
Let $(S, \Omega)$ be a $DC(0)$, (Ricc=0). If type 0 $\Rightarrow$ type X, then every X-extension is a 0-extension. (The types X under consideration must respect the dominant energy condition).
\end{theorem}

Physically, this means that if in an open subset $\Omega$ (admitting a Cauchy hypersurface), there is no fluid and no electromagnetism (that is $\Omega$ is a $DC(0)$), there won't be neither in any X-extension. One can say that, because of the dominant energy condition, no physical object can go faster than light and therefore no physical object can enter into an expansion domain of $\Omega$. We can in fact expect such results to be extendable to other types than type 0. For example, if in $\Omega$ there is only a fluid but no electromagnetism, can there be electromagnetism in an expansion domain of $\Omega$ ? (said differently, if type 1 $\Rightarrow$ type X, can we expect any X-extension to be a 1-extension ?)

 \subsection{Physical causality and Lorentzian Rigidity.}
 
 From a common practical point of view, a good physical theory is one such that, knowing a set of physical data at a given moment, one can, using this theory, predict the evolution of these data with time, with a sufficient precision. Within general relativity, if one consider only physical data defined using the geometry of the Lorentzian manifold, we can translate this by saying that a good physical theory is one that, for an observer that would know at a given moment $t_0$ of his proper time the metric tensor $g$ on a $DC(X)$ $(S, \Omega )$, would enable this observer to deduce without ambiguity (up to isometry) the metric $g$ on a maximal expansion domain of $(S,\Omega)$ and therefore that would enable him to predict the evolution of the physical data defined using the metric $g$ on an interval of his proper time  corresponding to the portion of his worldline included (starting at $C(t_0)$ ) in this domain of expansion. Obviously, there will be no ambiguity only with the condition that all the expansion domain of $(S,\Omega)$ are isometrical one to another. This is the case when one can apply theorem 2, that is when $(S,\Omega)$ is of type I, meaning $\Omega$ contains a fluid with no electromagnetism, or nothing. However, if $(S,\Omega)$ is of type II, theorem 3 says that, even if the observer knows $g$ on $\Omega$, he won't be able to predict anything (more precisely, he will have an infinity of possible choices) about the evolution of data defined with the metric  $g$, (meaning here those defined for electromagnetism of type 2,3,4). Electromagnetism defined only using the geometry of the manifold is insufficient for a "good" physical theory.

\emph{Remark : } Of course, if one introduces on the manifold an exact 2-form and imposes the energy-momentum tensor corresponding to the Maxwell equations (this is the classical presentation), one recover a "good" theory, as one can use the known Cauchy-Choquet-Bruhat theorems applied to these data.

For the less experimented reader, we remind here that in general relativity, an observer can only collect all the data of a causal domain  $(S,\Omega)$ if he is at the temporal limit of the domain of expansion of $(S,\Omega)$. Therefore, any prediction on the evolution of physical data can only be approximate. The approximation is usually made in the following manner : knowing the data on $(S,\Omega) $ at time $t_0$, the observer postulates "by extension" the data on a domain $(S',\Omega ')$ which contains strictly $(S,\Omega)$, ($S\subset S'$ and $\Omega \subset \Omega '$). He can then predict, via this extension, the evolution of the data on a time interval  such that his corresponding world line stays in the domain of expansion of $(S',\Omega ')$, in the case where this expansion is unique up to isomorphism. Think for example about the case of the use of Schwarzchild geometry : knowing that a star or a planet is spherically symmetric as well as some portion of space around it, the observer postulate that some larger domain is still spherically symmetric. He can then compute the geodesics of this larger domain to predict the movement of satellites ( doing then another approximation, that is the fact that these small bodies does not affect the global geometry, but this is more an approximation than a theoretical obstruction like the one we are talking about).

\newpage

\section{The three main concepts: types, Cauchy theorems, rigidity.}

We want to sum up here the main ideas of the above sections.

The Universe, or space-time, is a Lorentzian manifold, that we do not suppose to be of dimension 4, equipped with a metric of signature $(-, +, +, ... , +)$.  We introduce no "matter" nor "fields" in space-time,  matter and fields are only domains of the manifolds where the geometry satisfies some particular conditions. The equations of physics are only translation of identities or theorems of Riemannian geometry, like for example the second Bianchi identity , the computation of the divergence of a tensor using the curvature, etc...

Concerning determinism, to do physics is to be able to predict the results of experiments. Mathematically, we translate this by some rigidity property  of the domain of space-time we are studying, domain therefore equipped with a particular type, and to obtain a "good" physical theory on this domain means to prove a Cauchy theorem on it.

We therefore give the three main following definitions : We now consider Lorentzian manifolds $(M, g)$ or $(\Omega, g)$ of dimension $n\geqslant4$.

\begin{definition}
Let $(\Omega, g)$ be a Lorentzian manifold, or a domain of such a manifold. To define a \emph{type} on $(\Omega, g)$ is to define properties $P(g)$ that depend only on the geometry of $(\Omega, g)$, that is properties that are left invariant by any isometry of $\Omega$ : $P(\varphi^{*}g)=P(g)$ for any isometry $\varphi$ on $\Omega$. It is important to note that an isometry being also an homeomorphism and a diffeomorphism, $P(g)$ also depends on the topological and the differential structure of $\Omega$.
\end{definition}

We remind that : a \emph{causal} domain $(\Omega, S, g)$ is given by a time-oriented manifold $(\Omega, g)$ with a space like hypersurface $S$ that is a Cauchy hypersurface for $(\Omega,g)$. A causal domain of type $P(g)$ is a causal domain $(\Omega, S, g)$ where $(\Omega, g)$ is of type $P(g)$. Please review also the definitions of 4.2 on extensions of causal domains.

\begin{definition}
Let $(\Omega, S, g)$ be a causal domain of type $P(g)$. We will say that $(\Omega, S, g, P(g))$ is \emph{rigid}, or \emph{ satisfies the Cauchy property}, or also that it is \emph{deterministic}, if the maximal $P(g)$-extension domain of $(\Omega, S, g, P(g))$ is unique up to isometry.
\end{definition}

\begin{definition}
To prove a Cauchy theorem for a causal domain $(\Omega, S, g)$ of type $P(g)$ is to prove that $(\Omega, S, g, P(g))$ is rigid. This is where one must do (difficult) mathematics; see all the Choquet-Bruhat theorems.
\end{definition}

Let us precise more heuristically what we mean with the idea of type. A physical theory is a model for a region of space-time. Our point is that choosing a model for a region of space-time is to chose a type for this region; for example, a perfect fluid is a domain of type 1, a charged fluid is a domain of type 2, etc... To obtain a "good theory" is to prove a Cauchy theorem for the type one has chosen as a model. (May be one could also say that in quantum theory, one can have a good model, i.e. a good "type", without rigidity theorem...)

\newpage

\section{Five dimensional space-time.}
As we saw in section 3, the classical objects of electromagnetism cannot be obtained from the geometry of a 4-dimensional Lorentzian manifold. What can we try ? Obviously, if we want to obtain geometrically more objects, we need to enrich the geometry. Historically, one of the most famous method is due to Nordstrom, Kaluza and Klein : it consists in augmenting the dimension of the space-time manifold.

We shall see that starting with this idea and building upon the ideas presented in section 5, we will be able to propose a unified geometrical setting for both gravitation and electromagnetism. This will be obtained by suppressing a requirement usually made in papers on the subject, that is imposing a Ricci flat metric on the 5-dimensional space-time manifold, requirement which is not justified from our view point. This will be explained in subsection 6.3.

\subsection{"Small" dimension.}
How can we model a small fifth dimension ? The method originally proposed by Kaluza and Klein was to use a 5-dimensional fibre bundle structure over a 4-dimensional base representing classical space-time. We will see that a $S^1$-principal bundle will indeed lead to a model for space-time where gravitation and electromagnetism will be unified when not considering a Ricci-flat metric.

However we consider here that this structure is already too rigid. A fibre structure is a good model for the notion of direction, but the projection is a strong and rigid data. We want here to try to find a minimal extension of the model of general relativity.

The first minimal idea we can propose is the following :

\textbf{Hypothesis 1 : } Space-time is represented by a Lorentzian manifold $(M,g)$ of dimension 5 for which there exists a $\epsilon >0$ such that through any $x\in M$ there exists a unique closed loop, not homotopic to a point, and whose length is less than $\epsilon$. (The requirement of the existence of a small $\epsilon$ is not mathematically necessary, but models the notion of "small" dimension). 

This minimal idea seems however difficult to exploit mathematically and probably physically. It seems more reasonable to be closer to the idea of another dimension to ask for this small loop to be a totally geodesic submanifold. We therefore set :

\textbf{Hypothesis 2 :} Space-time is represented by a Lorentzian manifold $(M,g)$ of dimension 5 for which there exists a $\epsilon >0$ such that through any $x\in M$ there exists a unique spacelike submanifold $S_x$ of dimension 1, totally geodesic, compact, not homotopic to a point, and of diameter less than $\epsilon$ for the metric induced by $g$. We suppose furthermore that the submanifold field $x \mapsto S_x$ is differrentiable $(C^k)$,(meaning that for all $x\in M$, there exists a neighborhood $V_x$ and a $C^k-$vector field $Y$ on $V_x$ such that $\forall p \in V_x$, $Y_p$ is a basis for $T_pS_p$ (which means here $0 \neq Y_p \in T_pS_p$)).

Each manifold $S_x$ (supposed to be connected) is therefore diffeomorphic to a circle and is the image of a geodesic of $M$. Mathematically, we have defined a \emph{fibration} of $M$ by the submanifolds $S_x$. The main advantage of hypothesis 2 is that it gives a natural normalized vector field, unique up to orientation, that will, as you guess, represent the electromagnetic potential.

\subsection{Mathematical aspects}

From now on, we will adopt hypothesis 2, adding the fact that the manifold (or at least the domain we observe) is time-oriented.

\textbf{Hypothesis 2 :} Space-time is represented by a time-oriented Lorentzian manifold $(M,g)$ of dimension 5 for which there exists a $\epsilon >0$ such that through any $x\in M$ there passes a unique connected submanifold $S_x$ of dimension 1, spacelike, totally geodesic, compact, not homotopic to a point, and of diameter less than $\epsilon$ for the metric induced by $g$. We suppose furthermore that the submanifold field $x \mapsto S_x$ is differrentiable.

\begin{definition}
For all $x\in M$, choosing an orientation on $S_x$, there exists a neighborhood $V_x$ of $x$ in $M$ on which we can define a vector field $Y$ by setting, for each $p\in V_x$, $Y_p$ to be the tangent vector to $S_p$ at $p$ such that $g(Y_p,Y_p)=1$ and corresponding to the chosen orientation. We then define a 1-form $Y^*$ associated to $Y$ by the metric $g$ , ($Y^*=Y^b=Y_i$ if $Y=Y^i$). Then, we define the horizontal space $H_p$ for $p \in V_x$ as the subspace of $T_pM$ $g$-orthogonal to $Y_p$. At last, we note $F=d(Y^*)$ the differential of $Y^*$.
\end{definition}

Caution: because of the two possible orientation at each point, it is not always possible to define a field $Y$ continuous on all of $M$. The choice made has no incidence on the results that follow.

Some remarks. If $Y=Y^i \partial_i$ in a chart, then $Y^*=Y_idx^i$. Furthermore, the connexion being without torsion, we have $F=dY^*=\nabla_iY_j-\nabla_jY_i$.  We will denote, for all $x\in M$, $H_x=(Y_x)^{g\bot}$, the horizontal space in $x$. Having chosen a model simpler than the usual fibre-bundle one, we need a tensor field measuring the fact that $Y$ is not a Killing vector field.
\begin{definition}
We call $K$-factor, or Killing defect, the following tensor : 
$$K=K_{ij}= \nabla_i Y_j +\nabla_j Y_i$$
\end{definition}

If $K=0$, the flow generated by $Y$ is an isometry field, and it is then easy to turn $M$ into a circle principal bundle. We then recover the classical Kaluza-Klein setting. We will see that as a particular case in section 6.6.

We have defined all of our mathematical setting. We now are going to show that it gives, under a very natural definition of fluid, seen as a geometrical type-domain of space-time, using only geometrical theorems, and not posing any law, the Einstein-Maxwell-Lorentz equations as well as all the classical conservation laws.

\subsection{Do not kill Ricci.}
Hypothesis 2 produces a natural vector field $Y$, and from this, a 2-form $F$ that we identify with the electromagnetic field 2-form. This is now the point where we depart from the articles we have seen. In these, it is always considered that 5-dimensional space-time must be Ricci-flat. However, in the frame of Kaluza-Klein theory, this implies with the usual hypothesis made, that $|F|_g=0$, which contradicts the requirement of electromagnetism. Thus nothing can be obtained this way.

However, from our point of view, there is no reason to ask for the Ricci curvature to be zero. We can see this Ricci=0 requirement as a way to consider that matter is "added" to space-time; geometry cames next. From our point of view, there is only geometry, thus curvature ; matter is only an aspect of geometry.
 
Relieving this "Ricci=0" requirement, we will see that the known Kaluza-Klein formulae give the classical Einstein and Maxwell-Lorentz equations, that is gravity and electromagnetism, using only geometrical theorems or formulae. 

We will start by giving very general equations for space-time dynamics as seen by a family of timelike observers. We will then see that if these observers are linked to a massive charged fluid, defined in a purely geometrical way, and if they can only see what is happening on their 4-dimensional space-time, they will recover the classical equations of physics.

\subsection{Equations for space-time dynamics.}
\subsubsection{General equations}
We consider given, on an open subset of $M$ where $Y$ is defined, a vector field $X_0$, timelike, of norm $g(X_0,X_0)=-1$, and orthogonal at each point of $M$ to $Y$, $(X_0)_x \,\bot_g\, Y_x$. This vector field represents a family of observers.
 
We recall that $H_x=Y_x^{\bot}$. We define a subspace $H'_x$ of $H_x$ by $$H'_x=<X_0(x), Y_x>^{\bot}.$$ $H'_x$ can be interpreted as the space seen by $X_0(x)$.

We note $G_{ij}=R_{ij}-\frac{1}{2}Sg_{ij}$ the Einstein curvature. We set $^eG= \,^eG_i \, ^j$ the assoociated endomorphisms field. We note $^e G_H$ the endomorphisms field on the horizontal subspaces $H_x$ defined by $^eG_H=pr_H \circ (^eG_{|H})$, where for $x \in M$, $(pr_H)_{|x}$ is the orthogonal projection of $T_x M$ on $H_x$.

We define an endomorphisms field $^eG'$ by : $^eG'(X_0)=\,^eG'(Y)=0$ (in all point $x$), and $^eG'(X)=\, ^eG_H(X)$ if $X \in H'$.

To write down the matrix of $G$ and $^eG$, we then complete $X_0$ and $Y$ by three vector fields $X_1,X_2$ et $X_3$, not canonical, such that $(X_0,X_1,X_2,X_3, Y)$ is in every $x \in M$ a $g$-orthonormal basis of $T_x M$. Then $(X_0,X_1,X_2,X_3)$ is a basis for $H_x$, and $H'_x=<X_1,X_2,X_3>$. 

At last, we define two vector fields $Z$ et $Z'$ by
$$Z=pr_{H'} \circ  \,^eG(Y)$$
and $$Z'=pr_{H'} \circ \, ^eG(X_0).$$ 
We write:
$$ ^eG(X_0)=- \mu X_0 + Z' -e Y$$
and, as $G$ is symetric, $$ ^eG(Y)=e X_0+ Z + \gamma Y$$
where $\mu$ and $e$ are functions on $M$. We shall also write $Z=\sum_1^3 Z^i X_i$ and $Z'=\sum_1^3 Z'^i X_i$.

In the basis $(X_0,X_1,X_2,X_3, Y)$ for $T_x M$, the matrices of $G$ and $^eG$ are writen:
\[
G=\left(\begin{array}{ccc} \mu & \left(\begin{array}{ccc} Z'^1 & Z'^2 & Z'^3 \end{array}\right) & e 
\\  \left(\begin{array}{c} Z'^1 \\ Z'^2 \\ Z'^3 \end{array}\right)& (G')_{|H'} & \left(\begin{array}{c} Z^1 \\ Z^2 \\ Z^3 \end{array}\right) 
\\ e &  \left(\begin{array}{ccc} Z^1 & Z^2 & Z^3 \end{array}\right)  & \gamma \end{array}\right)
\]
and 
\[
^eG=\left(\begin{array}{ccc} -\mu & \left(\begin{array}{ccc} Z'^1 & Z'^2 & Z'^3 \end{array}\right) & e 
\\  \left(\begin{array}{c} Z'^1 \\ Z'^2 \\ Z'^3 \end{array}\right)& (^eG')_{|H'} & \left(\begin{array}{c} Z^1 \\ Z^2 \\ Z^3 \end{array}\right) 
\\ -e &  \left(\begin{array}{ccc} Z^1 & Z^2 & Z^3 \end{array}\right)  & \gamma \end{array}\right)
\]
In tensorial form, and using only the natural fields besides $X_0$, we can write, by noting again $G=G^{ij}$ the 2-times contravariant tensor associated to $G_{ij}$ by $g$:
 \begin{equation}
\fbox{ $\begin{split}
G= &\mu X_0 \otimes X_0 -e(X_0 \ts Y+Y \ts X_0)+\gamma Y \ts  Y + G' \\
       &+ (Z \ts Y+Y \ts Z) + (Z' \ts X_0+ X_0 \ts Z')
       \end{split}$}
\end{equation}
Please note that, being given $X_0$, the functions $\mu, \,e$ and $\gamma$ are natural as :

$-\mu=G(X_0,X_0)$, $e=G(Y, X_0)$, and $\gamma =G(Y,Y)$.\\

Here are some basic properties of the $g$-orthonormal basis $(X_0,X_1,X_2,X_3, Y)_x$ of $T_x M$.

\begin{proposition}
We note $(X,Z)$ the scalar product $g(X,Z)$.
\begin{enumerate}
  \item $D_Y Y=0$ because $Y$ is tangent to geodesics and normalized.
  \item $\forall p$, $(D_{X_p} Y, Y)=0$ as $(Y,Y)\equiv 1$.
 \item $(D_Y X_p, Y)=0$ as $(X_p,Y)\equiv 0$ and therefore $D_Y (X_p,Y)=0=(D_Y X_p,Y)+(X_p,D_Y Y)$.
  \item $(D_Y X_p, X_p)=0$ as $(D_YX_p, X_q)=-(X_p, D_Y X_q)$.
  \item for the same reason : $(D_{X_r} X_p, X_p)=0$
  \item $D_{X_0} Y= 1/2 (\,^eF(X_0)+ \, ^eK (X_0))$. (see below)
 \end{enumerate}
\end{proposition}
We noted $div_g T =cDT$ and $d^*T=-cDT$ for a 2-times contrariant tensor $T$, where $c$ is contraction; this definition is sufficient if $T$ is symetric (for example $G^{ij}$, or $R^{ij}$), otherwise, one has to say on which index contraction is made.

We are now going to show three equations, which can be understood as the equations for the dynamics of space-time as seen by the observers $X_0$.

First of all, by definition, $F=d Y^*$, and therefore $dF=0$. This is the first Maxwell equation.

Let us now compute $div_g F$ to find what will become, for a fluid, the second Maxwell equation. 

We will note $^eF$ the tensor $^eF=\nabla_i Y^j-\nabla^j Y_i$. In each point $x$, it is the endomorphism of $T_xM$ associated to $dY^* = \nabla_i Y_j-\nabla_j Y_i$ by $g$. In the same manner, we shall note $^eK=K_i \, ^j$ the endomorphism associated to $K$. We are in fact going to compute $div_g (^eF)$.

$^eF=\nabla_i Y^j-\nabla^j Y_i= 2\nabla_i Y^j -K_i \, ^j$.
\begin{align*}
div_g (^eF) & = \nabla^i (2\nabla_i Y^j-K_i \, ^j) \\
                     & = 2\nabla^i \nabla_i Y^j - \nabla ^i  K_i \, ^j
\end{align*}
but $\nabla _i Y^j =K_i \, ^j - \nabla^j Y_i$. Therefore
$div_g (^eF)= -2\nabla^i \nabla^j Y_i + \nabla ^i  K_i \, ^j$

Now, $\nabla^i\nabla^j Y_i =\nabla^j\nabla^i Y_i + R^j_l Y^l$. So :

$div_g (^eF)= -2\nabla^j \nabla^i Y_i -2R^j_l Y^l - \nabla ^i  K_i \, ^j$.

Then, $R^j_l = G^j _l +1/2S g^j_l$, and $$G^j_l Y^l= e X_0^j +\gamma Y^j+G'^j_lY^l+Z^j$$
but $G'^j_l Y^l=\, ^e G'(Y)=0$ by definition, and $G_l^j Y^l =\, ^eG(Y)= e X_0 +\gamma Y+Z$. Noticing that $g^j_l =id$, one obtain the important relation :
$$\fbox{$ R^j_l Y^l =e X_0 +\gamma Y+ Z +1/2 S_g . Y $} $$
$\nabla ^j \nabla^i Y_i = \nabla^j(divY)=grad(div Y)$, and $\nabla^i K_i^j= div_g K$.
We then obtain the "second Maxwell equation" :
\begin{equation}
\fbox{ $-div_g(^eF)= 2e. X_0 + 2[grad(div_g Y) +div_g K + Z ]+ (2 \gamma+S_g).Y $}
\end{equation}
Remark: $div_g Y= \frac{1}{2} tr_g K$. Indeed, $tr_g K= 2\nabla^i Y_i$. Therefore, if $K=0$, then $div_g Y=0$.

We are now going to apply the second Bianchi identity, $div_gG:=cD(G):=\nabla^iG_{ij} = 0$, to obtain an equation that we shall interpret as the Einstein-Lorentz equation, (movement of charged particles in gravitational and electromagnetic fields.)

Using the proposition above, we find using equation (4) : 
\begin{equation}
\fbox{ $\begin{split}
\nabla_i G^{ij}=& (D_{X_0}\mu )X_0 +\mu D_{X_0}X_0 + \mu(div_g X_0)X_0\\
                            & -(D_{X_0} e )Y-(D_Ye)X_0\\
                            &-e[(div_gX_0)Y+D_{X_0}Y+D_YX_0+(div_gY)X_0]\\
                            &+(\nabla_i (G')^{ij})+(D_Y \gamma)Y+\gamma(div_gY)Y\\
                            &+D_ZY+D_YZ+(div_gZ)Y+(div_gY)Z\\
                            &+D_{Z'}X_0+D_{X_0}Z'+(div_gZ')X_0+(div_gX_0)Z'\\
                            &=0
\end{split} \label{bg} $}
\end{equation}

Interpreting $\nabla_i G^{ij}$ as a vector, we project this relation on $X_0$, $Y$, and $H$, which means that we write successively : $g(\nabla_i G^{ij},X_0)=0$, $g(\nabla_i G^{ij},Y)=0$, and 
$pr_{H}(\nabla_i G^{ij})=0$. We obtain, noting again $(X,Z)$ the scalar product $g(X,Z)$:
\begin{equation}
\fbox{ $\begin{split}
g(\nabla_i G^{ij}, X_0)=& -div_g(\mu X_0)  -e(X_0, D_{X_0} Y)+e .div_gY +D_Y e\\
                            &+(X_0, \nabla_i (G')^{ij})+(X_0,D_ZY)+(X_0,D_YZ)\\
                            &+(X_0,D_{Z'}X_0)+(X_0,D_{X_0}Z')-div_gZ'\\
                            &=0
 \end{split} $}
 \end{equation}
 where we used $D_{X_0}\mu +\mu. div_gX_0=div_g(\mu X_0)$.
 
Then
\begin{equation}
\fbox{ $\begin{split}
g(\nabla_i G^{ij},Y)=&-div_g(eX_0)+\mu (Y,D_{X_0}X_0)+(Y,\nabla^i G'_{ij})+div_g(\gamma Y)\\
                            &+div_gZ+(Y,D_ZY)+(Y,D_YZ)\\
                            &+(Y,D_{X_0}Z')+(Y,D_{Z'}X_0)\\
                            &=0
 \end{split} $}
 \end{equation}
At last, projecting (\ref{bg}) on $H$, one obtains :
\begin{equation*}
\begin{split}
&pr_{H}\{ \mu D_{X_0}X_0 -e[D_{X_0}+D_Y X_0] + (\nabla_i G'^{ij})\\
&+D_ZY+D_YZ+(div_gY)Z+D_{Z'}X_0+D_{X_0}Z'+(div_gX_0)Z' \} \\
&=0
\end{split}
\end{equation*}
One then remark the following important relation: $^eF=\nabla_i Y^j-\nabla^j Y_i$ and $^eK=\nabla_i Y^j+\nabla^j Y_i$. Therefore $2DY=\,^eF+\,^eK$, and $$\fbox{$D_{X_0}Y=\frac{1}{2}[^eF(X_0)+\,^eK(X_0)]$}$$
Writing now $D_YX_0=D_{X_0}Y-[X_0,Y]$, the last projection can be written  :
\begin{equation}
\fbox{ $\begin{split}
pr_{H}&\{ \mu D_{X_0}X_0 -e.[^eF(X_0)+\,^eK(X_0)]+ (\nabla_i G'^{ij})\\
&+e[X_0,Y]+D_ZY+D_YZ+(div_gY)Z\\
&+D_{Z'}X_0+D_{X_0}Z'+(div_gX_0)Z' \} \\
&=0
\end{split}$}
\end{equation}
This last equation can be interpreted as the equation of the dynamic measured by the family of observers $X_0$ that could only see what happens on $H$, their "perceived" space-time, and not being able to see what happens on the fifth dimension carried by $Y$. But we shall also see, and we think that this is quite remarkable, that the two other projections, on $X_0$ and $Y$, correspond exactly to the conservation laws in the case of a very natural definition of a charged fluid. Please note that changing $Y$ in $-Y$ lead to changing $e$ by $-e$.

To end this section, one can mimic some classical definitions : 
\begin{enumerate}
\item An observer of $(M,g)$ is a timelike curve $\gamma : I \rightarrow M$.
\item The space-time seen by $\gamma$ at $x=\gamma (t)$ is $Y_x^{\bot}$.
\item The full space seen by $\gamma$ at $x=\gamma (t)$ is $\gamma ' (t) ^{\bot}$.
\item The classical space seen by $\gamma$ at $x=\gamma (t)$ is $<Y_{\gamma (t)}, \gamma '(t)>^{\bot}$, (i.e $H'_{\gamma (t)}$). 
\item A  \emph{classical}, or \emph{galilean}, observer, is a timelike curve $\gamma$  which is horizontal, i.e. $\gamma '(t) \bot Y_{\gamma (t)}$ for all $t$. If such an observer can only see 4 dimensions and not the fifth carried by $Y$, then his "measure process" are projections on his horizontal space-time, $H_{\gamma (t)}$.
\end{enumerate}

\subsubsection{Perfect fluids}
We now apply these equations to matter and electromagnetism.

Let us remind that we denote $^e G_H$ the endomorphisms field on the horizontal spaces $H_x=(Y_x)^{\bot}$ defined by $^eG_H=pr_H \circ (^eG_{|H})$, where for $x \in M$, $(pr_H)_{|x}$ is the orthogonal projections of $T_x M$ on $H_x$. Building up on classical models, we define:

 \emph{A fluid is a domain of $M$ where there exists a naturally defined timelike vector field. More precisely, at least to begin with, a fluid is a domain of $M$ where $^eG_H$ admits in each point a eigenspace of dimension 1, timelike, and orthogonal to $Y$.} 
 
 The equations of the previous section are then valid for $X_0$, an eigenvector of this subspace, unitary and in the orientation We are going to show that the classical equations for a matter fluid appear as an approximation, where according to our point of view approximation must be understood as "additional geometric hypothesis".

We begin by setting:
\begin{definition}
(Domain of type 2, charged matter fluid.). Let $\Omega \subset M$ be a domain of the manifold $(M,g)$. We say that $\Omega$ is a type 2 domain, or a charged matter fluid domain, if the Einstein curvature $G$ of $\Omega$ satisfies:
\begin{enumerate}
\item $\forall x \in \Omega$, $^eG_H(x)$ has an eigenvalue -$\mu (x) <0$ whose eigenspace $E_{-\mu}(x)$ is of dimension 1 and timelike. 
\item $\forall x \in \Omega$, $^eG_x(Y_x) \in <E_{-\mu}(x),Y_x>$, the subspace generated by $E_{-\mu}(x)$ and $Y_x$.
 \end{enumerate}
 \end{definition}

In a type 2 domain, we define\emph{ naturally} a unit vector field $X_0$, timeline, in the orientation, and such that $E_{-\mu} (x) =<X_0 (x)>$. Then, with the notation of the previous section, hypothesis 1 means that $Z'=0$, and hypothesis 2 means that $Z=0$. Note then that in any point $x$, the subspace $<X_0(x),Y_x>$ is stable by $^eG_x$.  

As in the previous section, to write down the matrices, we add to $X_0$ and $Y$ three vector fields $X_1,X_2$ et $X_3$, not canonical, such that $(X_0,X_1,X_2, X_3, Y)$ is in any point $x \in M$ a $g$-orthonormal basis for $T_x M$. Then $(X_0,X_1,X_2,X_3)$ is a basis for $H_x$, and $(X_1,X_2,X_3)$ is then a basis for the natural subspace $H'_x=<X_0(x),Y_x>^{\bot}$ of $H_x$. $H'_x$ can be seen as the space "perceived " by the observer $(X_0)_x$ linked to the fluid. We then keep all the notation of the previous section.

The matrices of   $G$ et $^eG$ can then be written  :
\[
G=\left(\begin{array}{ccc} \mu & \begin{array}{ccc}  0&0&0 \end{array} & e 
\\  \begin{array}{c} 0\\0\\0  \end{array}& (G')_{|H'} & \begin{array}{c} 0\\0\\0 \end{array} 
\\ e &  \begin{array}{ccc} 0&0&0 \end{array}  & \gamma \end{array}\right)
\]
and 
\[
^eG=\left(\begin{array}{ccc} -\mu & \begin{array}{ccc}  0&0&0 \end{array} & e 
\\  \begin{array}{c} 0\\0\\0  \end{array}& (G')_{|H'} & \begin{array}{c} 0\\0\\0 \end{array} 
\\ -e &  \begin{array}{ccc} 0&0&0 \end{array}  & \gamma \end{array}\right)
\]
In tensorial form, we can write :
 \begin{equation}
\fbox{ $
G= \mu X_0 \otimes X_0 -e(X_0 \ts Y+Y \ts X_0)+\gamma Y \ts  Y + G' $}
\end{equation}

The dynamics equations of the previous section then become:

-Second Maxwell equation:
\begin{equation}
\fbox{ $-div_g(^eF)= 2e. X_0 + 2[grad(div_g Y) +div_g K ]+ (2 \gamma+S_g).Y $}
\end{equation}

-Baryonic number conservation law:
\begin{equation}
\fbox{ $\begin{split}
g(\nabla_i G^{ij}, X_0)=& -div_g(\mu X_0) -e(X_0, D_{X_0} Y)+e .div_gY \\
                            &+(X_0, \nabla_i (G')^{ij})\\
                            &=0
 \end{split} $}
 \end{equation}

-Charge conservation law:
\begin{equation}
\fbox{ $\begin{split}
g(\nabla_i G^{ij},Y)&=-div_g(eX_0)+\mu (Y,D_{X_0}X_0)+(Y,\nabla^i G'_{ij})+div_g(\gamma Y)\\
                            &=0
 \end{split} $}
 \end{equation}

And at last the motion equation as seen by the observers $X_0$ linked to the fluid :
\begin{equation}
\fbox{ $pr_{H}\{ \mu D_{X_0}X_0 -e.[^eF(X_0)+\,^eK(X_0)]+ (\nabla_i G'^{ij})+e.[X_0,Y]\}=0$}
\end{equation}

To justify the names we have given to these equations, let us give a more sophisticated model of fluid, where by this we mean a model with more geometrical hypothesis:
\begin{definition}
(Type 2' domain, perfect fluid). Let $\Omega \subset M$ be a domaine of the manifold $(M,g)$. We say that $\Omega$ is a type 2' domain, or a perfect fluid, if $\Omega$ is a type 2 domain on which, furthermore, $K=0$, i.e., $Y$ is a Killing vector field.
\end{definition}
The hypothesis that $Y$ is a Killing vector field will be a very important hypothesis in the rest of the paper. It means that $Y$ generates a local group of isometries. We have seen that this also implies that $div_gY=0$. It also implies that $D_Y \gamma=D_Y e=D_Y \mu=0$ as $\gamma$, $e$ and $\mu$ are intrinsic characteristics of $G$ and as it implies that $[X_0,Y]=-\mathcal{L}_Y X_0=0$, as can be proved. 

At this point, it is interesting to note that we can in fact give a much more natural (when compared to classical general relativity) definition for a fluid.

\begin{proposition}
 A domain $\Omega \subset M$ is a	perfect charged matter fluid domain if and only if its Einstein curvature tensor can be written 
 $$G=\mu X \otimes X + \alpha Y \otimes Y+P$$
with the condition that, at each point $x$, $pr_H(X)$ is a basis for a timelike 1-dimensional eigenspace of $^eG_H$ of eigenvalue $-\mu<0$, and $P$ is a matrix such that $^eP(X)=\, ^eP(Y)=0$. If $P=0$ (which means a fluid with no pressure), then $X$ is unique for the decomposition $G=\mu X \otimes X + \alpha Y \otimes Y$.

For such a perfect fluid, there is a unique decomposition (once a time orientation is chosen) 
$G= \mu X_0 \otimes X_0 -e(X_0 \ts Y+Y \ts X_0)+\gamma Y \ts  Y+G'$, 
where $g(X_0,X_0) =-1$ and $X_0 \bot Y$, and satisfying all the conditions of type 2 (and of the paragraph that follows its definition). $\mu$ is called mass density, $e$ the charge density. These are canonically given by : $\mu=-G(X_0,X_0)$, $e=G(Y,X_0)$, $\gamma= G(Y,Y)=|F|_g+1/2S_g$.
 \end{proposition}

The previous equations then get much simpler forms:

-SecondMaxwell equation:
 $$-div_g(^eF)= 2e. X_0 + (2 \gamma+S_g).Y $$
 
-Baryonic number conservation law:
$$g(\nabla_i G^{ij}, X_0)= -div_g(\mu X_0) -e(X_0, D_{X_0} Y)+(X_0, \nabla_i G'^{ij})=0$$

-Charge conservation law:
$$g(\nabla_i G^{ij},Y)=-div_g(eX_0)+\mu (Y,D_{X_0}X_0)+(Y,\nabla^i G'_{ij})=0$$
But, because $K=0$, $^eF=\nabla_i Y^j-\nabla^j Y_i= 2\nabla_i Y^j -K_i \, ^j=2\nabla_i Y^j=2DY$. $^eF$ being anti-symmetric, and $G'$ being symmetric, developing $0=\nabla_i (G'(Y))$ shows that $(Y,\nabla^i G'_{ij})=0$.

Concerning the motion equation as seen by the observers $X_0$ linked to the fluid:
$$pr_{H}\{ \mu D_{X_0}X_0 -e.^eF(X_0)+ (\nabla_i G'^{ij})+e.[X_0,Y]\}=0$$

But the hypothesis $K=0$, implying the fact that $Y$ generates isometries, also implies that $[X_0,Y]=-\mathcal{L}_Y X_0=0$ and thus that $D_{X_0}X_0$ is horizontal, and that $ (Y,D_{X_0}X_0)=-(X_0, D_{X_0} Y)=0$. Indeed, we therefore have $D_YX_0=D_{X_0}Y$, and we have seen in the proposition of the previous subsection that $(D_YX_0,X_0)=0$. Therefore $0=([X_0,Y],X_0)=(D_{X_0}Y,X_0)-(D_YX_0,X_0)=-(Y,D_{X_0}X_0)$ which means that $D_{X_0}X_0$ is horizontal.

The equations therefore become:

-Maxwell equations : $dF=0$, as $F=dY^*$, and
\begin{equation}
\fbox{ $-div_g(^eF)= 2e. X_0 + (2 \gamma+S_g).Y $}
\end{equation}

-Baryonic number conservation law:
\begin{equation}
\fbox{ $g(\nabla_i G^{ij}, X_0)= -div_g(\mu X_0) +(X_0, \nabla_i G'^{ij})=0$}
 \end{equation}

-Charge conservation law:
\begin{equation}
\fbox{ $g(\nabla_i G^{ij},Y)=-div_g(eX_0)=0$}
 \end{equation}

And at last the motion equation as seen by the observers $X_0$ linked to the fluid:
\begin{equation}
\fbox{ $ \mu D_{X_0}X_0 -e.\,^eF(X_0)+ (\nabla_i G'^{ij})=0 $}
\end{equation}

That is, we have recovered the classical equations of physics ! We then call of course, $e$ the charge density, $\mu$ the Baryionic density, and $G'^{ij}$ the pressure of the fluid. We then get exactly the classical Einstein-Maxwell-Lorentz equations. Note that the projection of the second Maxwell equation on $H$, representing what can be "seen" by the observers $X_0$, is exactly $-div_g(^eF)= 2e. X_0$. Furthermore, if we consider a fluid without pressure, i.e. $G'=0$, the conservation laws become $div_g(\mu X_0)=div_g(eX_0)=0$, and the motion equation $\mu D_{X_0}X_0 =e.\,^eF(X_0)$, which is exactly Lorentz law. At last, if we choose $G'=\lambda .Id$, we recover exactly the perfect fluids equations. We think it is quite remarkable that the conservation laws are obtained naturally, and not as \emph{ad hoc} hypothesis. 

\emph{Remark :} The factor 2 in front of $e.X_0$ in the second Maxwell equation is just a matter of convention. Replace the einstein tensor $G$ in the given definitions of fluid by $\T G= 2\eta^{-1}.G$ to get $-div_g(^eF)= \eta .e. X_0 + (\eta . \gamma+S_g).Y $, the other equations being unchanged. See the computation of $div_g(^eF)$ above.

Note also that, when $K=0$, that is for a type 2' fluid, the coefficient $\gamma$ takes a clear meaning:  $\gamma = |F|_g+1/2S_g$.

We think that it is important to understand that the type 2 and 2' fluids models are just approximate general fluids models. Indeed, one recovers the last equations, identical to the classical ones,  if in the equations of the previous subsection one considers that the projections  on $H'$ and $<X_0>$ of $Z, Z', Y$ and of their derivatives $DZ, DZ', DY$ can be "neglected" when compared to the other expressions for the Riemannian measure induced on $H'$ and $<X_0>$ by $g$.

\subsubsection{Free fall in an electrogravitational field.}
One of the cornerstone of general relativity is the equivalence principle. It is expressed mathematically by the hypothesis that free particles follow time-like geodesics of space-time, or, for perfect fluids without pressure (dust), whose energy-momentum tensor is $\mu X_0\otimes X_0$, by the fact that the vector field $X_0$ associated to the flow lines is a geodesic vector field. We recalled at the beginning of this paper that this fact is obtained by applying the Bianchi identity to this tensor when considered as the Einstein curvature; once again it is just a purely geometrical fact.

Considering the inclusion of electromagnetism in the geometrical frame of space-time, it would be satisfactory to extend this principle to our five-dimensional setting. We considered a domain $\Omega$ of space-time to be a charged perfect fluid of matter without pressure if its curvature tensor is 
 \begin{equation}
\fbox{ $
G= \mu X_0 \otimes X_0 -e(X_0 \ts Y+Y \ts X_0)+\gamma Y \ts  Y$}
\end{equation}
where $Y$ is geodesic and Killing. In this case, the equation of movement on classical four-dimensional space-time $Y^{\bot}=H$ is $\mu D_{X_0}X_0 =e.\,^eF(X_0)$; $X_0$ is not of course geodesic. However, we would like it to be the "trace" on classical space-time, that is the projection on $H$, of a geodesic trajectory in five dimensions. Obviously this has to involve movement along the "small" fifth dimension. We have the following thorem:
\begin{theorem}
With the above hypothesis, (perfect charged fluid without pressure), the vector field 
$$X:= X_0-\frac{e}{\mu}Y$$
is a geodesic vector field (considering of course $\mu \neq 0$ everywhere).
\end{theorem}
To prove this, one can just compute $D_X X$, the only thing to notice being that $X_0(\frac{e}{\mu})=0$. 

However, we prefer to show how we were lead to this simple but convincing result. As we said, a geodesic movement would have to imply the fifth dimension, and we want its projection on $H$ to be the flow of $X_0$. It was therefore natural to look for a vector field of the form $W=X_0+ \alpha Y$ for some function $\alpha$. Computing $D_WW$, we get :
\begin{align*}
D_{X_0+\alpha Y}(X_0+\alpha Y) & = D_{X_0} X_0+(X_0(\alpha)+\alpha Y(\alpha) ).Y+\alpha (D_{X_0} Y+D_Y X_0)+\alpha ^2 D_Y Y\\
                     & =D_{X_0} X_0+(X_0(\alpha)+\alpha Y(\alpha) ).Y+\alpha (D_{X_0} Y+D_Y X_0)
\end{align*}
because $Y$ is geodesic. We now use a few facts : first, the connection is torsion-free, and $Y$ is killing. Therefore $[X_0,Y]=0$, and thus $D_{X_0}Y=D_Y X_0$. Secondly, $2D_{X_0}Y=\, ^eF(X_0)$, as $K=0$. And thirdly, $D_{X_0}X_0=\frac{e}{\mu}. ^eF(X_0)$. So
$$D_{X_0+\alpha Y}(X_0+\alpha Y)=(\frac{e}{\mu}+\alpha )\, ^eF(X_0)+ (X_0(\alpha)+\alpha Y(\alpha)).Y$$
But we have seen that $^eF(X_0)$ is horizontal, that is, orthogonal to $Y$. Therefore, $D_W W$ will be zero if $\frac{e}{\mu}=-\alpha$ and $X_0(\alpha)+\alpha Y(\alpha)=0$. We thus have to check that 
$$X_0 (\frac{e}{\mu})-\frac{e}{\mu}.Y(\frac{e}{\mu})=0.$$
$Y$ generating isometries, $Y(\frac{e}{\mu})=0$. So it remains to check that $X_0 (\frac{e}{\mu})=0$.

This conservation law is a simple consequence of the two we already obtained: $div_g(\mu X_0)=div_g(e X_0)=0$. Indeed, developing these equations gives:
$$\mu .div_gX_0+X_0(\mu)=0$$
$$e.div_gX_0+X_0(e)=0$$
Multiplying the first equation by $e$ and the second by $\mu$, then subtracting,  gives $X_0 (\frac{e}{\mu})=0$.
\\

Let us sum up our results. The starting point is :
\begin{theorem}
(Einstein, 1916) Space-time is a 4-dimensional Lorentzian manifold $(M,g)$. A perfect dust fluid is a domain $\Omega$ of space-time whose Einstein curvature is of the form $G=\mu X_0 \otimes X_0$, $X_0$ being a timelike vector field. The Bianchi identity implies that $X_0$ is a geodesic vector field, and that $div_g(\mu X_0)=0$. To modelize electromagnetism, one then add a closed 2-form $F$, a function $e: \Omega \rightarrow \mathbb{R}$, and postulate the Lorentz and second Maxwell equations, as well as the conservation of charge $div_g (eX_0)=0$. (In fact, given $F$ and $e$, it is sufficient to postulate the first and second Maxwell equations, Bianchi giving the Lorentz law.)
\end{theorem}

Our results in this section are the following:
\\ 

\textbf{Setting :} Space-time is a 5-dimensional Lorentzian manifold $(M,g)$ satisfying hypothesis 2 of section 6.2. The vector field $Y$ and the closed 2-form $F=dY^{*}$ are then naturally defined. We now suppose that the vector field $Y$, beside being geodesic, is also a Killing vector field. 

Let us remind that, if $G$ is the Einstein curvature tensor, we note $^e G_H$ the endomorphisms field on the horizontal spaces $H_x=(Y_x)^{\bot}$ defined by $^eG_H=pr_H \circ (^eG_{|H})$, where for $x \in M$, $(pr_H)_{|x}$ is the $g$-orthogonal projection of $T_x M$ on $H_x$. 

\emph{A fluid of matter is a domain $\Omega$ of $M$ where, at each point,  $^eG_H$ has a timelike 1-dimensional eigenspace.}

\begin{definition} 
 A domain $\Omega \subset M$ is a	perfect charged matter fluid domain if and only if at each point :  $^eG_H$ has a timelike 1-dimensional eigenspace $E_{-\mu}$ of eigenvalue $-\mu<0$, and $^eG(Y) \in<Y,E_{-\mu}>$. 
 
 This is the case if and only if its Einstein curvature tensor can be written 
 $$G=\mu X \otimes X + \alpha Y \otimes Y+P$$
with the condition that, at each point $x$, $pr_H(X)$ is a basis for a timelike 1-dimensional eigenspace of $^eG_H$ of eigenvalue $-\mu<0$, and $P$ is a matrix such that $^eP(X)=\, ^eP(Y)=0$. 

If $P=0$ (which means a fluid with no pressure), then $X$ is unique for the decomposition $G=\mu X \otimes X + \alpha Y \otimes Y$ (once a time orientation is chosen).
 \end{definition}

\textbf{Associated "classical" data : } For such a perfect fluid without pressure, there is a unique decomposition
$G= \mu X_0 \otimes X_0 -e(X_0 \ts Y+Y \ts X_0)+\gamma Y \ts  Y$, 
where $g(X_0,X_0)=-1$ and $X_0 \bot Y$. $\mu$ is called mass density, $e$ the charge density. These are canonically given by : $\mu=-G(X_0,X_0)$, $e=G(Y,X_0)$, $\gamma= G(Y,Y)=|F|_g+1/2S_g$. We then have $X=X_0-\frac{e}{\mu}Y$.
\\

Summing the computations of the previous subsection, our result is then :
\begin{theorem}
For a perfect charged fluid without pressure $G=\mu X \otimes X + \alpha Y \otimes Y$, and its associated classical data as above, Bianchi identity gives:
\begin{itemize}
\item Conservation Laws: $X_0(\frac{e}{\mu})=div_g(\mu X_0)=div_g(eX_0)=0$ 
\item Maxwell equations :  $dF=0$ and $-div_g( ^eF)=2eX_0+(2\gamma +S_g)Y$
\item Free Fall : $X=X_0-\frac{e}{\mu}Y$ is a geodesic vector field.
\item Lorentz equation: $\mu D_{X_0} X_0=e. ^eF(X_0)$. This is just free fall read on $H$.
\end{itemize}
When projected on the "classical" 4-dimensional space-time $H=Y^\bot$, these equations give the classical equations of physics. 
\end{theorem}

Note that Lorentz equation is obtained from the geodesic motion of $X$ by developing $D_X X=0$ and writing this equation on the horizontal space $H=Y^{\bot}$. Comparing this with the proof of theorem 5, we therefore see that : 

 \emph{Free fall for $X$ is equivalent to Lorentz equation for $X_0$}. 
 
 (Remember that the first Maxwell equation, $dF=0$, is always obvious as we set $F=dY^*$.)

A perfect charged fluid with pressure is a domain whose curvature is of the form $$G=\mu X \otimes X + 
\alpha Y \otimes Y+P$$
for some matrix $P$ such that $^eP(X_0)=\, ^eP(Y)=0$. $P$ is called the pressure/constraint tensor. An analogous theorem can be obtained from the results of the previous subsection, also giving the equations of classical physics in the presence of pressure. The choice of $P$ corresponds to the choice of a state equation for the fluid :
$$-div_g(^eF)= 2e. X_0 + (2 \gamma+S_g).Y $$
$$div_g(\mu X_0) -(X_0, \nabla_i P^{ij})=0$$
$$div_g(eX_0)=0$$
$$ \mu D_{X_0}X_0 -e.\,^eF(X_0)+ (\nabla_i P^{ij})=0 $$

\subsection{Other geometrical-physical equations.}

The above equations were obtained using identities of Riemannian geometry . But there exist other such equations. One can then ask wether we could obtain other equations having a physical meaning.

For example, one can compute $tr_g G$, which by definition is $-3/2S_g$, but also $-\mu+tr_g G'+\gamma$, from where we get another equation : $$\mu-tr_gG'-\gamma=3/2S_g$$

These kind of equations already exist in classical 4-dimensional general relativity.

There is also another very important equation that wasn't used in this paper, the Raychaudhuri equation  (see Hawking-Ellis, Wald, Choquet-Bruhat). It gives the evolution of a family of geodesics defined as the flow lines of a vector field $X$ satisfying $D_XX=0$. This equation, usually given in a 4-dimensional space-time, is valid in any dimension, and it gives important motion equations. It is a purely geometrical equation, lying on basic Riemannian geometry identities.

\subsection{A particular case: Kaluza-Klein theory.}
It is very important to note that until now, in this section, we have never mentioned any kind of energy-momentum tensor. The point is that from our point of view, this concept has no meaning.

However, to make this paper complete, we are going to show that with our point of view, that is without requiring $Ricci=0$, we can recover the usual energy-momentum tensors  from the classical Kaluza-Klein theory, which now appears as a particular case of the models of the previous subsection.
It is also important to remark that this enable to avoid the Lagrangian methods, which are always slightly ambiguous when applied to fluids. 

The goal of this subsection is only to provide a link with the classical works on Kaluza-Klein theory. It is therefore too long when compared to the previous subsections, and it might be a good idea for the reader to proceed now to section 7, where extension of our five-dimensional model to higher dimensions is given for inclusion of other possible physical effects, and come back here when interested.

\subsubsection{Mathematical aspects.}
We present directly the modern description of Kaluza-Klein theory. 

\emph{The model for space-time is a 5-dimensional Lorentzian manifold $(M,g)$ equipped with a principal $S^1$-fibre bundle structure, ( $S^1$ being the circle).}

Principal fibre bundle theory can be found in several textbooks, e.g. Kobayashi-Nomizu . However, in the case of a $S^1$-bundle, the theory is much simpler; we therefore here present an elementary vision. A very important reference for us is  [Bourguignon].

\begin{definition} (Hypothesis:) We consider a Lorentzian manifold $(M,g)$ of dimension 5, time-oriented, $g$ being of signature $(-,+,+,+,+)$, such that the Lie group $S^1$ acts transitively and freely on $M$ ; therefore $\B{M}:=M/S^1$ is a manifold. Furthermore, there exists a Riemannian submersion $\pi : (M,g) \rightarrow (\B{M},\B{g})$ such that $\B{g}$ has the signature $(-,+,+,+)$ and such that $\forall x \in \B{M}$, $\pi^{-1}(x)$ is spacelike. We also suppose in this subsection that $vol_g \pi^{-1} (x)=cst$ on $M$. One then shows that the fibers $\pi^{-1}(x)$ are geodesics of $M$.
\end{definition}
We therefore see that $M$ appears as a particular case of the manifolds studied in the previous subsection with hypothesis 2. One just needs to set $K=0$, to ask for the submanifold $S_x$ to be the fiber $\pi^{-1}(x)$, and to set $vol_g (\pi^{-1}(x))=\epsilon$.

Starting from here, we can define natural objects :
\begin{definition}
By choosing an orientation, we define a vector field $Y$ on $M$, by setting that in each point $x$ of $M$, $Y_x$ is the vector tangent to the fiber $\pi^{-1}(\pi(x))$ at $x$ and such that $g(Y,Y)=1$. We then define a 1-form $Y^*$ associated to $Y$ by $g$, ($Y^*=Y^b=Y_i$ if $Y=Y^i$). We then define the horizontal space $H_x$ at $x\in M$ as being the subspace of $T_x (M)$ $g$-orthogonal to $Y_x$. We can then write, slightly abusing notations, $g=\B{g} + Y^*\otimes Y^*$. At last, we note $F=d(Y^*)$ the differential of $Y^*$.
\end{definition}
Remark: thanks to the operation of $S^1$, a continuous choice can be made for the orientation of $Y$.

We recall some notations. We note $G_{ij}=R_{ij}-\frac{1}{2}Sg_{ij}$ the Einstein curvature. $^eG_i^j$ is the associated endomorphism. We note $^e G_H$ the endomorphisms field on the horizontal spaces $H_x$ defined by $^eG_H=pr_H \circ (^eG_{|H})$, where for $x \in M$, $(pr_H)_{|x}$ is the orthogonal projection of $T_x M$ on $H_x$.

It is easy to show that because of the $S^1$-invariance of $F$, we can define unambiguously a 2-form $\B{F} \in \Lambda^2(\B M)$ such that $\pi^* \B{F} = F$.

To obtain the relation between the curvature of $M$ and of $\B{M}$, it is interesting to set particular basis of the tangent spaces. Writing with a bar all the objects of $\B M$, let's fix some more notations. For any tangent vector $X \in T_x M$, we denote $X^h$ the orthogonal projection on $H_x$ and $X^v$ the one on $<Y_x>$, the subspace generated by $Y_x$. $d\pi_x$ is an isometry between $H_x$ and $T_{\B{x}}\B M$ where $\B{x}=\pi(x)$. We then denote, for a vector $\B{X}$ of $T_{\B{x}}\B M$, $\B X^r$ the horizontal pullback of $\B X$, i.e. $\B X^r \in H_x$, and $d \pi_x(\B X^r)= \B X$.

Using the fibre bundle structure and the isometry $d\pi_x$, we now build the following basis field. We choose a horizontal vector field $X_0$ (i.e $\in H_x$ for all $x$), timelike on $M$, everywhere orthogonal to $Y$, and such that $g(X_0,X_0)=-1$. We then add to $X_0$ vectors forming a $g$-orthonormal basis $(X_0,X_1,X_2,X_3)$ of $H_x$. The fields $X_p$ can be built in such a way that there exist vector fields $\B X_p$ on $\B M$ forming a basis of $T_{\B{x}}\B M$ in each point, and such that $(\B X_p)^r = X_p$. 

Every thing then lies on the following formulae, due to O' Neill, and that can be found in [Bourguignon].
\begin{proposition}
For vector fields $X,Z$ on $M$, associated to vector fields $\B X, \B Z$ on $\B M$ such that $\B  X^r = X$ et $ \B Z^r =Z$, we have:
\begin{itemize}
  \item $R(X,Z)= \B R (\B X, \B Z) + 2 \B F \circ \B F (\B X, \B Z)$. 
  \item $R(X,Y)=-2d^* F(X)=2div_g F(X)$.
  \item $R(Y,Y)=  |F|^2_g$
  \item $S=\B S - |F|^2_g$
\end{itemize}
\end{proposition}

$\B F \circ \B F (\B X, \B Z)$ must be understood as meaning $\B F_i \,^l \B F_{lj}X^i Z^j$. Besides, we denoted $div_g T =cDT$ and $d^*T=-cDT$ for a tensor field $T$. Caution: the notation $ |F|^2_g$ is misleading; it means $ |F|^2_g=F^{ij}F_{ij}$ and therefore, because of the signature of $g$, it can be $\leq0$.

Easy computations show that : $G_{ij}=R_{ij}-1/2(\B S -1/4 |F|^2_g)g_{ij}$.

We then sum up the equations of the previous proposition under matrices form:
\[
R_{IJ}=\left(\begin{array}{cc}
 
  \B R_{ij}+ \B F_i \,^l \B F_{lj} & \begin{array}{c} 1/2(div_g F)_0 \\ \vdots \\ 1/2(div_g F)_3\end{array}  \\
 \begin{array}{ccc} 1/2(div_g F)_0 & \hdots & 1/2(div_g F)_3 \end{array} & 1/4 |F|^2_g
\end{array}\right)
\]
and
\[
G_{IJ}=\left(\begin{array}{cc}
 \B G_{ij}+1/2( \B F_i \,^l \B F_{lj} +1/4 |F|^2_g.g_{ij}) &
  \begin{array}{c} 1/2(div_g F)_0 \\ \vdots \\ 1/2(div_g F)_3\end{array} \\
 \begin{array}{ccc} 1/2(div_g F)_0 & \hdots & 1/2(div_g F)_3 \end{array} &  3/8 |F|^2_g-1/2S
\end{array}\right)
\]
If we define, (just out of luck !), $T^F_{ij}=-1/2( \B F_i \,^l \B F_{lj} +1/4 |F|^2_g.g_{ij})$, the last matrix can be written :
\[
G_{IJ}=\left(\begin{array}{cc}
 \B G_{ij}-T^F_{ij}&  \begin{array}{c} 1/2(div_g F)_0 \\ \vdots \\ 1/2(div_g F)_3\end{array} \\
 \begin{array}{ccc} 1/2(div_g F)_0 & \hdots & 1/2(div_g F)_3 \end{array} & 3/8 |F|^2_g-1/2S
\end{array}\right)
\]
Then :
\[
^eG_I\,^J=\left(\begin{array}{cc}
 \B G_i\,^j-(T^F)_i\,^j& \begin{array}{c} 1/2(div_g F)_0 \\ \vdots \\ 1/2(div_g F)_3\end{array} \\
\begin{array}{ccc} -1/2(div_g F)_0 & \hdots & 1/2(div_g F)_3 \end{array}&\gamma
\end{array}\right)
\]
where $\gamma=3/8 |F|^2_g-1/2S$. This is where appears the remarkable idea of Kaluza and Klein producing naturally the electromagnetic tensor $T^F$.

We then remark that we have:
$$^eG(Y)=1/2(div_g F)_0.X_0+ \hdots + 1/2(div_g F)_3.X_3+ \gamma.Y$$
and therefore :
\begin{equation} pr_H(^eG(Y))=pr_H(1/2div_g F)\end{equation}

Here are the basic properties of the $g$-orthonormal basis $(X_0,X_1,X_2,X_3, Y)_x$ of $T_x M$, that can be deduced from the analogous proposition of subsection 6.4.1.

\begin{proposition}
We note $(X,Z)$ the scalar product $g(X,Z)$.
\begin{enumerate}
  \item $D_Y Y=0$
  \item $\forall p$, $D_Y X_p=D_{X_p} Y$, i.e. $[X_p,Y]=0$.
  \item $D_Y X_p$ et $D_{X_p} Y$ are horizontal.
  \item $(D_Y X_p, X_p)=(D_{X_p} Y,X_p)=0$
  \item $(D_{X_r} X_p, X_p)=0$
  \item $D_{X_p} X_p$ is horizontal.
  \item $div_g Y =0$.
  \item The endomorphism $^eF= \,^e(dY^*)$ associated to $F$ by $(^eF)_i \, ^j = F_i \, ^j$ satisfies $^eF= 2DY$. So, $^eF(X_0)=2D_{X_0} Y$.
  \item $div_g F =2 R_{ij} Y^j$
\end{enumerate}
\end{proposition}

We now have settled all the tools and mathematical definition needed for what follows...

\subsubsection{Kaluza-Klein electromagnetism.}
We saw that in 4 dimensions, we were unable to define geometrically the objects of classical electromagnetism, essentially because there was no way to produce naturally a corresponding 2-form; we only had the null-trace part of $G_{ij}$ corresponding to $T^F$.

 We saw that starting with the choice of a \emph{type} modeling a classical charged fluid, but now in a 5-dimensional space-time, we could define \emph{geometrically} all the objects of classical physics and furthermore obtain  the classical Maxwell-Lorentz equations for an observer who would only "perceive" 4 dimensions. In the frame of Kaluza-Kein theory, we choose this model by "pulling back" using $\pi$ the 4-dimensional model of a charged fluid. The matrices of $G$ now naturally produce the electromagnetic tensor $T^F$. To get closer to classical known physical results, we add a few geometrical requirements, which means, once again, doing some physical approximations. 

\begin{definition}
(2K type domain, charged fluid). Let $\Omega \subset M$ be a domain of the manifold $(M,g)$. We say that $\Omega$ is a 2K type domain (2K for Kaluza-Klein), if the Einstein curvature $G$ of $\Omega$ satisfies:
\begin{enumerate}
  \item $\forall x \in \Omega$, $^eG_H$ admits : 
  \begin{itemize}
  \item An eigenvalue -$\mu (x) <0$ whose eigenspace $E_{-\mu}(x)$ is timelike and of dimension 1. 
  \item A second eignvalue $\lambda (x)$, $-\mu(x)<\lambda(x)<\mu(x)$, such that dim$E_{\lambda}(x)$=3 and $E_{\lambda}\, _g\bot \,E_{-\mu}$.
 
\end{itemize}
  \item $\forall x \in \Omega$, the subspace $<E_{-\mu}(x), Y(x)>_{T_x M}$ is stable under $^eG$. 
  
\end{enumerate}
\end{definition}

In a 2K type domain, we define naturally \emph{naturally} a unitary vector field, timelike and in the orientation, and such that $E_{-\mu} (x) =<X_0 (x)>$. We then choose a basis $X_1, X_2, X_3$ of $E_\lambda$ such that $(X_0, X_1, X_2, X_3, Y)$ is a $g$-orthonormal basis of $T_x M$. In this basis, the matrices of $G$ and $^eG$ are the following:
\[
G_{IJ}=\left(\begin{array}{ccccc} \mu & 0 & 0 & 0 & e \\0 & \lambda & 0 & 0 & 0 \\0 & 0 & \lambda  & 0 & 0   \\0 & 0 & 0 & \lambda & 0 \\ e & 0 & 0 & 0 & \gamma \end{array}\right)\,and\, \,\,
^eG_I\,^J=\left(\begin{array}{ccccc} -\mu & 0 & 0 & 0 & e \\0 & \lambda & 0 & 0 & 0 \\0 & 0 & \lambda  & 0 & 0   \\0 & 0 & 0 & \lambda & 0 \\ -e & 0 & 0 & 0 & \gamma \end{array}\right)
\]
Indeed, because of the stability hypothesis, $^eG(X_0)=aX_0-eY$. So, $G_H(X_0)=pr_H (aX_0-eY)=aX_0$, thus $a=-\mu$. For the same reason, $^eG(Y)=a' X_0+\gamma Y$, and by symmetry, $a'=e$.

Remark : $e$ est "naturally" given by $e= -G(Y,X_0)$. Also, $\gamma =G(Y,Y)$.

We therefore immediately see that $$pr_H(div_g F)=2 eX_0$$ which is Maxwell second equation.
The first Maxwell equation, $ dF=0$, is obvious, as $F$ is exact and therefore closed, $F=dY^*$.

Here again, the choice of a \emph{geometric} model, a type corresponding to the natural notion of fluid, gives very easily the two Maxwell equations. We then apply the results of subsection 6.4.2 to recover the classical motion equation as well as the conservation laws. (We can also recover them slightly more easily using the proposition of the previous subsection). We therefore get:

Maxwell equations : $dF=0$, as $F=dY^*$, and
\begin{equation}
\fbox{ $-div_g(^eF)= 2e. X_0 $}
\end{equation}

-Baryonic number conservation law:
\begin{equation}
\fbox{ $-g(\nabla_i G^{ij}, X_0)= div_g(\mu X_0) +\lambda div_gX_0=0$}
 \end{equation}

-Electrical charge conservation law:
\begin{equation}
\fbox{ $-g(\nabla_i G^{ij},Y)=div_g(eX_0)=0$}
 \end{equation}

And at last the motion equation as seen by the observers $X_0$ linked to the fluid:
\begin{equation}
\fbox{ $ (\mu+\lambda) D_{X_0}X_0 -e.\,^eF(X_0)+ grad_{1,2,3}\lambda=0 $}
\end{equation}

Pulling back using $\pi$ the types 3 or 4 seen in section 3.2 on the dimension 4, we also recover all the classical results in a purely geometrical way.

For any vector field $\B X, \B Y$ on $\B M$, and their pull backs $X:= \B X ^r$ and $Y:=\B Y ^r$, we have $D_X Y=(D_{\B X} \B Y)^r + 1/2 [X,Y]^v$ and $div_{\B g} \B X \circ \pi = div_g X$. Therefore, if we suppose that the observers $X_0$ can only see or measure, (c.f. next section ), what happens on $\B M$, we see once again that by projecting using $\pi$ the above formulae, we recover exactly the classical Einstein-Maxwell-Lorentz equations for the family of observers  $\B X_0$.

\subsection{Remark on physical measurement in five dimensions.}
The equations obtained for a family of observers are general; they follow from the sole hypothesis of a fifth geodesic dimension. A fluid of matter is a choice of a geometric setting. Once a family of observers is defined, the equations obtained with these hypothesis can be seen as measures made by the observers. As such, and considering they can only "see" four dimensions, they can be made in two ways. Either these measures "neglect" everything happening on the fifth dimension, and this corresponds to projecting all the equations on $H$ or $H'$. Or the measures consists in taking the "mean" value along the fifth dimension, and this corresponds to going to the quotient, which is exactly the frame of Kaluza-Klein theory. In both cases, one recover the classical physical equations.

\subsection{The cosmological constant.}
If needed, one can easily introduce the cosmological constant $\Lambda$ in our model by calling $\Omega$ a type $2\Lambda$ domain if $G+\Lambda g$ satisfies the hypothesis of type 2 (or 2').

\subsection{Timelike fifth dimension and metric signature.}
We have chosen a spacelike fifth dimension as this is usually what is done. However, for the results of this paper, there is no need to do so. Indeed, choosing the fifth dimension to be timelike, that is choosing the metric $g$ to be of signature $(-,+,+,+,-)$, introduces no change in the results and formulae obtained here. Essentially, one just need to replace the charge density $e$ by $-e$.

Definitions of fluid must be slightly modified to require that the 1-dimensional timelike eigenspace of $^eG$ should have a timelike $g$-orthonal projection on $Y^{\bot}$, (the fluid is not "flowing along the fifth dimension").

In fact, results obtained by the first author, Michel Vaugon, towards a geometrical frame for quantum mechanics in the spirit of this paper, indicate that it might be useful, or even necessary, to consider such a signature for the metric. See the related paper indicated in section 7.5.

\newpage

\section{Beyond five dimensions.}
Theoretical evidences, like string theory, suggest the need for more than five dimensions. We want to present in this section a possible extension of our model, that preserve the results obtained until now for the inclusion of electromagnetism, but that enable the possible inclusion of such other dimensions that might model geometrically other physical effects. Note however that we do not pretend here to model precisely such other fields, but that we just want to present a possible mathematical frame.

\subsection{Minimal extension}
In this subsection, we do not want to present directly the most general definition, but instead to present the evolution of the authors ideas leading to the most general geometric setting which will include the preceding ones, including the above 5-dimensional "electrogravitational" space-time, as particular cases. This subsection can be skipped with no influence on the rest of the paper.

We can propose the following minimal definition, extending our previous definition of section 6.1: 

\begin{definition}
 $M$ is a Lorentzian manifold of dimension $n=4+m$ for which there exists $\epsilon>0$ such that in any point $x \in M$, there exists a submanifold $S_x$, of dimension $m$, totally geodesic and not contractible (i.e not having the homotopy type of a point), space like, and whose diameter (for the metric induced by $g$) is less than $\epsilon$. We suppose that the submanifold field $x \mapsto S_x$ is differentiable, meaning that in some neighborhood of any $x\in M$ there exist $m$ vector fields forming a basis for $T_pS_p$ for all $p$ in this neighborhood.
\end{definition}

We could also imagine that the signature of $g$ is of any kind, $S_x$ not being necessarily spacelike.

We want to show here that our model for gravitation and electromagnetism does not exclude the possibility for more than 5 dimensions. Many models can be tried, we are not looking for maximal generality here; the reader can try others...

\begin{definition}
$M$ is a Lorentzian manifold of dimension $n=4+1+m$, for which there exists $\epsilon , \epsilon ' >0$ such that through any  $x \in M$, there exists a unique submanifold $V_x$ spacelike, compact, not contractible, totally geodesic, of dimension $m+1$ and whose diameter is $\leq \epsilon$, and such that furthemore:

-i/ there exists a unique submanifold of dimension 1, $\gamma _x$, of  $V_x$, passing through $x$, totally geodesic, compact and not homotopic to a point, whose diameter is that of $V_x$, and

-ii/ there exists a unique submanifold $W_x$ of $V_x$, of dimension $m$ and whose diameter is $\leq \epsilon '. \epsilon$ ($W_x \bot _g \gamma _x$ ?)

(As another possibility, we can replace ii/ by : any geodesic of $V_x$ closed and of diameter $< \epsilon$, is of diameter $\leq \epsilon '. \epsilon$.)
\end{definition}

The idea is of course that the dimensions "above" 5 are even smaller than the fifth. An example of such a manifold can be obtained by choosing $M$ to be a fibre bundle, whose fibers are $S^1\times W$, equipped with an adequate metric.

A fluide of type 2K would then be a domain of $M$ in which $G$ is of the following type : 

We still define $Y_x$ as the normalized tangent vector to $\gamma _x$. We decompose $T_x M$ as a $g$-orthogonal sum $T_xM= <X_0>\oplus H'_x\oplus <Y_x> \oplus E_x$, where $E_x$ is the tangent space of $W_x$ and where we keep some notation of the previous sections. We then set $U_k=pr_{|E_x}\circ \, ^eG(X_k)$ for $k=0,1,2,3$ and $U_4=pr_{|E_x}\circ \, ^eG(Y)$, where $(X_k)$ is a basis of $H'_x$. $G$ can then be written : 
 \begin{equation}
\fbox{ $\begin{split}
G= &\mu X_0 \otimes X_0 -e(X_0 \ts Y+Y \ts X_0)+\gamma Y \ts  Y + G' \\
       &+\sum_0^3 (U_k \ts X_k+X_k \ts U_k ) + (U_4 \ts Y+ Y \ts U_4)+G''
       \end{split}$}
\end{equation}

Just as the end of section 6.4.2, if we suppose that in the case of a charged matter fluid, we observe a type where the projection on $H'$, $<X_0>$ and $<Y>$ of the $U_k$ and of $G''$, as well as the projections of their derivatives $D U_k$ and $D G''$, can be neglected for the measures induced by $g$ on $H'$, $<X_0>$ and $<Y>$, we recover the Einstein-Maxwell-Lorentz equations of section 6.4.2.

This minimal extension thus seems reasonable. However, it requires the fact that all the "extra" dimensions are spacelike, and imposes the fact that the "electromagnetic" fifth dimension is larger than the other "small" dimensions. We want now to relieve these constraints. We will first use the usual fibre bundle structure, but then see that this structure can be obtained in a more natural, although equivalent, manner.

We therefore now want to extend the natural definition of a fluid given in definition 15 and the theorem 7 that follows to the case of 5+m dimensions. Of course, we write 5+m because we would like to preserve the results obtained for electromagnetism. The idea is course to set a geometrical frame for any other physical effect beyond macroscopic gravitation and electromagnetism.

We consider that our 5-dimensional model for gravitation and electromagnetism is a fair model; it will therefore be our starting point. We will use the classical structure of a fiber bundle to add extra dimensions. We will see in section 7.2 a more physical way to present this structure. We now suppose that the signature "along the fifth dimension" can be negative; we will note $\sigma=\pm 1$ this signature. For easiness of lecture we remind all the definitions in this subsection, just adding this possible new signature.

We thus start with a pseudo-riemannian manifold $(M,g)$ of dimension 5, $g$ being of signature $(-,+,+,+,\sigma)$, satisfying hypothesis 2 of section 6.2, adding the fact that the restriction of $g$ to each $S_x$ is of signature $\sigma$. Passing directly to the fiber bundle structure, we therefore have the following setting : 

\begin{definition} (Hypothesis:) We consider a Lorentzian manifold $(M,g)$ of dimension 5, $g$ being of signature $(-,+,+,+,\sigma )$, such that the Lie group $S^1$ acts transitively and freely on $M$ ; therefore $\B{M}:=M/S^1$ is a manifold. We suppose that for every $x \in M$, the orbit $S_x$ of $x$ under the action of $S^1$ is spacelike if $\sigma=+1$, timelike if $\sigma=-1$. Furthermore, there exists a Riemannian submersion $\pi : (M,g) \rightarrow (\B{M},\B{g})$ such that $\B{g}$ has the signature $(-,+,+,+)$ and such that $\forall x \in \B{M}$, $\pi^{-1}(x)=S_x$ We also suppose that $vol_g \pi^{-1} (x)=cst$ on $M$. One then shows that the fibers $\pi^{-1}(x)$ are geodesics of $M$.
\end{definition}

\begin{definition}
By choosing an orientation, we define a vector field $Y$ on $M$, by setting that in each point $x$ of $M$, $Y_x$ is the vector tangent to the fiber $\pi^{-1}(\pi(x))$ at $x$ and such that $g(Y,Y)=\sigma$. We then define a 1-form $Y^*$ associated to $Y$ by $g$, ($Y^*=Y^b=Y_i$ if $Y=Y^i$). We then define the horizontal space $H_x$ at $x\in M$ as being the subspace of $T_x (M)$ $g$-orthogonal to $Y_x$. We can then write, slightly abusing notations, $g=\B{g} + Y^*\otimes Y^*$. At last, we note $F=d(Y^*)$ the differential of $Y^*$.
\end{definition}
Remark: thanks to the operation of $S^1$, a continuous choice can be made for the orientation of $Y$.

Let us remind that, if $G$ is the Einstein curvature tensor, we note $^e G_H$ the endomorphisms field on the horizontal spaces $H_x=(Y_x)^{\bot}$ defined by $^eG_H=pr_H \circ (^eG_{|H})$.

It is easy to show that because of the $S^1$-invariance of $F$, we can define unambiguously a 2-form $\B{F} \in \Lambda^2(\B M)$ such that $\pi^* \B{F} = F$.
\\

Let us now consider a compact manifold $W$ of dimension $m$ equipped with a Riemannian metric $h$ of signature $(+,... ,+)$. The first natural way to add $m$ extra dimensions to our space-time model is to use a fibre-bundle structure:

\emph{Space-time is a $W$-fibre bundle whose total space is a 5+m-dimensional pseudo-riemannian manifold $(\T M, \T g)$, $\T g$ being of signature $(-,+,+,+,\sigma,+,...,+)$ and whose base space is the 5-dimensional manifold $(M,g)$ defined above.}

We therefore have a "double" fibration $\T M \rightarrow M \rightarrow \B M=M/S^1$. 

To make a simple image of that, we can heuristically say : $\B M$ is classical 4-dimensional space-time, $M$ is the "electrogravitational" 5-dimensional universe of section 6, and $\T M$ is the "real" 5+m dimensional universe.

\subsection{Observation atlas and fiber bundles.}
If we consider what we just described above as our first attempt to extend our model to more than 5 dimensions, we see that the main point is to define precisely additional "small", possibly "compactified", dimensions. The use of the fiber bundle structure is the natural way to proceed from a mathematician point of view. However, the use of the base space can seem, from a more "physical" point of view, to be useless. We therefore want to give here an alternative definition of a fiber bundle which insists on the fact that it is really a structure imposed on the total space and not on the base space, giving in particular a natural and intrinsic notion of "small dimensions" as fibers. This will be more appropriate for the physical use we have for this structure, as we consider the total space to be the "real" universe, and the base space only an approximate model, limited to restricted possible physical measurements.

\emph{Let be given a manifold $M$ of dimension $n=p+m$, and a compact manifold $W$ of dimension $m$. Consider a domain $D$ of $M$.}
\begin{definition}
An observation diffeomorphism on $D$ relative to $W$, or for short, a $W$-chart on $M$ is a triple $(\mathcal{V}, \phi, \Theta)$ where $\mathcal{V}$ is an open subset of $D$, $\Theta$ an open subset of $\mathbb{R}^p$, and $\phi$ a diffeomorphism from $\mathcal{V}$ onto $\Theta \times W$.
\end{definition}
\begin{definition}
A (observation) $W$-atlas on $D \subset M$ is a family $((\mathcal{V}_i, \phi _i, \Theta_i))_{i \in I}$ of $W$-charts such that :
\begin{itemize}
\item $\bigcup_{i \in I} \mathcal{V}_i = D$
\item $\forall i,j \in I$, $\forall x \in \mathcal{V}_i\cap \mathcal{V}_j$, $\phi^{-1}_i (\{ \phi_i^1(x)\} \times W)=\phi^{-1}_j (\{ \phi_j^1(x)\} \times W)$, where $\phi_i^1$ is the component function of $\phi$ going to $\Theta_i$.
\end{itemize}
\end{definition}

It is quite easy to prove that this definition is in fact equivalent to that of a fibre bundle :

\begin{theorem}
With the notations above, the two following propositions are equivalent:
\begin{itemize}
\item There exists an observation  $W$-atlas on $D$.
\item There exist a smooth manifold $B$ of dimension $n-m$ and a submersion $\pi : D \rightarrow B$ such that $(D, \pi , B)$ is a fibre bundle with fibre $W$.
\end{itemize}
\end{theorem}

Remark: the compacity of $W$ is required to prove that $B$ is Hausdorf. If not, one needs to require in definition 28 that any $\phi^{-1}_i (\{ \phi_i^1(x)\} \times W)$ is closed in $D$.
\\

We now want to consider the case where $W$ is a product manifold (for example $W=S^1 \times W'$). We therefore consider two compact manifolds $W_a$ and $W_b$.
\begin{definition}
For manifolds $D$ and $W=W_a \times W_b$ as above, an observation $W_a$-atlas is a family $((\mathcal{V}_i, \phi _i, \Theta_i))_{i \in I}$ of $W$-charts such that :
\begin{itemize}
\item $\bigcup_{i \in I} \mathcal{V}_i = D$
\item $\forall i,j \in I$, $\forall x \in \mathcal{V}_i\cap \mathcal{V}_j$, 
$$\phi^{-1}_i (\{ \phi_i^1(x)\} \times W_a \times \{ \phi_i^b(x)\} )=\phi^{-1}_j (\{ \phi_j^1(x)\} \times W_a \times \{ \phi_j^b(x)\} )$$
 where $\phi_i^1$ is the component function of $\phi$ going to $\Theta_i$ and $\phi_i^b$ the component going to $W_b$.
\end{itemize}

We define in the same way a $W_b$-atlas.

If $W_a$ is oriented, we say that the $W_a$-atlas preserves the orientation of $W_a$ if $\forall i,j \in I$, $\forall x \in \mathcal{V}_i\cap \mathcal{V}_j$, the orientation on $\phi^{-1}_i (\{ \phi_i^1(x)\} \times W_a \times \{ \phi_i^b(x)\} )$ carried from $W_a$ by $\phi_i$ is the same as that carried by $\phi_j$. (One write the same definition if $W_b$ is oriented).
\end{definition}

\begin{definition}
An observation $(W_a,W_b)$-atlas on $D$ is a family $((\mathcal{V}_i, \phi _i, \Theta_i))_{i \in I}$ of $W$-charts giving both a $W_a$-atlas and a $W_b$-atlas.
\end{definition}

One can then show :

\begin{proposition}
An observation $(W_a,W_b)$-atlas on $D$ is equivalent to a double fibration $$D \rightarrow \B D := D/W_a \rightarrow \B D /W_b$$
where $W_a$ and $W_b$ can be exchanged in the order of the fibration.
\end{proposition}

Caution : a $(W_a,W_b)$-atlas is not equivalent to a $(W_a \times W_b)$-atlas.

\newpage

\begin{definition} If $\mathcal{A}$ is a $(W_a,W_b)$-atlas, for every $x \in D$ and any $W$-chart $(\mathcal{V}_i, \phi _i)$ around $x$, we can define, \emph{naturally} relatively to $\mathcal{A}$, fibers through $x$ by 
$$W_{a,x} :=\phi^{-1}_i (\{ \phi_i^1(x)\} \times W_a \times \{ \phi_i^b(x)\} ) $$
and
$$W_{b,x} :=\phi^{-1}_i (\{ \phi_i^1(x)\}  \times \{ \phi_i^a(x)\} \times W_b) $$
which are respectively isomorphic to $W_a$ and $W_b$ and, this is the important point, these fibers do not depend on the chosen $W$-chart. If $W_a$ is oriented, so are all the fibers $W_{a,x}$.

Remark: if $x'\in W_{a,x}$, then $W_{a,x'}=W_{a,x}$.
\end{definition}

Caution : With a $(W_a,W_b)$-atlas, not being a $(W_a \times W_b)$-atlas, there is no naturally defined fiber $W_x$ isomorphic to $W_a \times W_b$.
\\

It is now of course interesting to consider the $W$-fibre bundle space-time $(\T M, \T g)$ of the previous subsection 7.1. as a manifold $\T M$ equipped with a $(S^1,W)$-atlas. According to the proposition above, we recover the same double fibration structure:
$$\T M \rightarrow M:= \T M /W \rightarrow \B M=M/S^1$$

All these definitions did not imply any metric. It is however now slightly easier than in the setting of a fibre bundle to fix the signatures on the fibers as defined above:

\emph{Given a manifold $M$ equipped with a $(W_a,W_b)$-altlas, a metric $g$ is compatible with the atlas if the signature of the restriction of $g$ to any fiber as defined above is constant. That is, for any $x \in M$, the signature of $g$ restricted to $W_{a,x}$ and $W_{b,x}$ is independent of $x$. In this case, for given signatures $\sigma_a$ and $\sigma_b$ of adequate length, we will say that $g$ is of signature $\sigma_a$ on $W_a$ and $\sigma_b$ on $W_b$. The signature of $g_x$ on the subspace of $T_x M$ orthogonal to $T_x(W_{a,x} \times W_{b,x})$ is then also independent of $x$; we call this subspace the horizontal space at $x$}.
\\

When this is done, suitable hypothesis on the metric $g$ can be made to turn the fibrations $D \rightarrow \B D := D/W_a \rightarrow \B D /W_b$ into semi-Riemannian submersions for adequate quotient metrics. However this will not be the case for a general compatible metric.
\\

There is a natural decomposition $T_x M= H_x \bot T_x W_{a,x} \oplus T_x W_{b,x}$. Note that  $T_x W_{a,x}$ is not necassarily orthogonal to $T_x W_{b,x}$.

$H_x$ will be called the apparent space-time at $x$. Caution : the field of tangent space $H_x$ need not be integrable ; there is not in general, even locally, a 4-dimensional submanifold of $M$ tangent at every $x$ to  $H$.

To conclude this section, we recall basic definitions concerning complete atlases. 

1/: Two observation $(W_a,W_b)$-atlases, possibly each compatible with a metric $g$, are said to be equivalent, it their union is still an observation $(W_a,W_b)$-atlas, compatible with $g$.

2/: The completed $(W_a,W_b)$-atlas associated to a $(W_a,W_b)$-atlas $A$, is the union of all the $(W_a,W_b)$-atlases compatible with $A$.

3/: A $(W_a,W_b)$-atlas is complete if it is equal to its completed $(W_a,W_b)$-atlas.

\subsection{Final model for 5+m-dimensional space-time. Matter fluid, visible and hidden pressure/constraint tensors. Main theorems for the motion of a general fluid.}
We now propose our final model for space-time. As we will see, considering two timelike dimensions leads to a very aesthetic model, preserving what we obtained for the unification of gravitation and electromagnetism in 5 dimensions. (Work by Michel Vaugon also suggests that two timelike dimensions might produce a very nice "differential geometric" setting for quantum mechanics, in the spirit of this paper; this is a story under writting...) The reader uncomfortable with this additional time, can still consider the signature on $S^1$ to be spacelike; only minor sign changes will be required in front of expressions using $e$ or $Y$, but all what follows remains in fact essentially unchanged.

\begin{definition}
Space-time is a $5+m$-dimensional semi-Riemannian manifold $(M,g)$ equipped with a complete $(S^1,W)$-atlas. We suppose that $g$ is compatible with this atlas. $g$ is of (total) signature $(-,+,+,+,-, +,...,+)$, of signature $-1$ on $S^1$, and of signature $(+,...,+)$ on $W$. The horizontal space at $x$ is $H_x := T_x(S_x^1 \times W_x)^{\bot}$ and $g_x$ is then of signature $(-,+,+,+)$ on $H_x$.
\end{definition}
As already noticed, the $(S^1,W)$-atlas on $M$ is equivalent to a "differential geometric" double fibration $M \rightarrow \B M:=M /W \rightarrow \B{\B M}:=\B M/S^1$, but these are not, in general, Riemannian submersion.
The effect of the "extra" $m$ dimensions carried by $W$ will be modeled via the geometry of the submersion $\pi: M \rightarrow \B M$ or, directly, via the metric $ g$ or its Einstein curvature $G$. Once again, the idea is that what passes to the quotient can be neglected.

\begin{definition} 
We define $Y$ to be the vector field defined at each $x \in M$ to be tangent to the fiber $S^1_x$ and such that $g (Y,Y)=-1$, with a chosen orientation for $S^1$. Again, $F=d(Y^*)$. We always suppose from now on that $Y$ is a geodesic and Killing vector field. (Note once again that if $Y$ is Killing and of constant norm, it is necessarily geodesic.) The local diffeomorphisms generated by $Y$ are therefore isometries.
\end{definition}

We suppose that we have the canonical standard orientation on $S^1$. We also add to the definition of $g$ being compatible with the given $(S^1,W)$-atlas on $M$ the following requirement : $\forall i,j \in I$ and observation-charts $\phi_i$, $\phi_j$, $\forall x \in \mathcal{V}_i\cap \mathcal{V}_j$, 
$\phi_i^*(\partial_t)_x$ and $\phi_j^*(\partial_t)_x$ are timelike and in the same time orientation, i.e. $g(\phi_i^*(\partial_t)_x,\phi_j^*(\partial_t)_x)<0$, where $\partial_t$ is the tangent vector to the canonical coordinates $(t,x,y,z)$ on $\Theta_i \subset \mathbb{R}^4$. This condition gives a "classical" time-orientation on every apparent space-time $H_x$, varying differentially with $x$.

\newpage

\subsubsection{General Fluids}
The idea we want to keep for the definition of a fluid is the fact that there is a natural timelike flow, or vector field, in classical space-time $\B {\B M}$. Considering now our basic "space-time" model as being the $5+m$-dimensional manifold $M$, this horizontal space, representing "classical" 4-dimensional space-time, is $H_x := T_x(S_x^1 \times W_{b,x})^{\bot}$. For a fluid, the important object will be once again the endomorphism field $^e  G_{ H}=pr_{H} \circ (^e  G_{|  H})$, which is essentially the endomorphism field $^e  G$, $g$-associated to the Einstein curvature $G$, restricted to the horizontal space $H$.

We start with the most general definition for a matter fluid : 

\begin{definition}
 A domain $D$ of $M$ is a fluid domain if, at every $x \in D$, the endomorphism $^e G_H=pr_H \circ (^e G_{|H})$ has a timelike 1-dimensional eigenspace $E_{-\mu}$ of eigenvalue $-\mu <0$, where we now consider $G$ as being twice the Einstein curvature tensor: 
 
 \center{\fbox{$G=2.Ricci_g -Scal_g.g:=2.Ric_g-S_g.g.$}}
\end{definition}

We now define \emph{naturally} the following objects:
\begin{itemize}
\item The vector field $X_0$, already seen, such that at every $x\in D$, $X_0(x)$ is the unique vector of the eigenspace $E_{-\mu}(x)$ in the chosen orientation, and such that $g(X_0,X_0)=-1$, which can be proven easily to be unique. The vector field $X_0$ will be called the \emph{apparent, or visible,} field of the fluid, and the associated flow, the \emph{apparent, or visible,} flow.
\item The smooth function $\mu :D \rightarrow \mathbb{R}$ defined by $\mu(x)=\mu_x$ where $-\mu_x$ is the eigenvalue associated to the eigenspace $E_{-\mu}$. It will be called the \emph{energy density} of the fluid.
\item The smooth function $e :D \rightarrow \mathbb{R}$ defined by $e(x)=G_x(X_0(x),Y_x)$. It will be called the electric charge density of the fluid.
\item The vector field $X=X_0+ \frac{e}{\mu}Y$, timelike, is called the vector field of the fluid, and the associated flow, the flow of the fluid.
\item The time-plane $\mathrm{T}_x$, of dimension 2, is the subspace of $T_xM$ generated by $X_0(x)$ and $Y_x$. Because $Y$ is Killing, it can be shown that $[X_0,Y]=0$, therefore the plane field $\mathrm{T}$ is integrable.
\item The time-tube $\mathcal{T}_x$ is then, at each $x\in D$, the integral submanifold passing through $x$ of the field $\mathrm{T}$. $\mathcal{T}_x$ is a submanifold of dimension 2, totally timelike. These "tubes" can be seen as a generalization of the flow lines of a fluid. They are oriented by the orientation of $X_0$ and $Y$. 
\end{itemize}

With these definitions, for all $x \in M$, the tensor $G$ restricted to $\mathrm{T}_x$ can be written  
$$G_{|\mathrm{T}_x}=\mu X \ts X+ \beta Y\ts Y$$ where $\beta :D \rightarrow \mathbb{R}$ is a smooth function.

Then, the tensor field $P=G-G_{|\mathrm{T}_x}$ will be called the fluid pressure. It satisfies $P(X_0,X_0)=P(X_0,Y)=P(Y,Y)=P(X,X)=0$. Therefore $G$ can be written :

 \begin{equation*}
\fbox{ $\begin{split}
G &=\mu X \ts X+ \beta Y\ts Y+P\\
       & =\mu X_0 \ts X_0+ e(X_0 \ts Y+Y \ts X_0)+(\beta+\frac{e^2}{\mu}). Y \ts  Y+P
       \end{split}$}
\end{equation*}

\begin{itemize}
\item The \emph{apparent pressure} $P_v$ will be the pressure $P$ restricted to the horizontal space $H_x$. That is : $\forall Z,Z' \in H_x$, $P_v(Z,Z')=P(Z,Z')$, and $\forall Z \in T_xD$, $\forall Z' \in T_xS^1 \oplus T_xW_x$, $P_v(Z,Z')=0$ and $P_v(Z,X_0)=0$.

\item The \emph{hidden pressure} is $P_h:=P-P_v$.
\end{itemize}

$G$ can now be written : \fbox{$G= \mu X \ts X+ \beta Y\ts Y+P_v+P_h$}
\\

In matrix form, with suitable basis for $T_xH$, $T_xS^1_x$ and $T_x W_x$, and with some abuse of notation for $P_h$, $G$ can be written: 

\[
G=\left(\begin{array}{cccc} \mu & \begin{array}{ccc}  0&0&0 \end{array} & e 
\\  \begin{array}{c} 0\\0\\0  \end{array}& P_v & \begin{array}{c} a\\b\\c \end{array} & P_h
\\ e &  \begin{array}{ccc} a&b&c \end{array}  & \gamma 
\\  & P_h & & P_h
\end{array}\right)
\]
where $a,b,c$ are the component of $P_h(Y)$ on $H'  := (T_x(S_x^1 \times W_{b,x}) \times <X_{0_x}>)^{\bot}$.
\\

The fluid will be called perfect if $P_h(Y)=0$. $G$ can then be written: 

\[
G=\left(\begin{array}{cccc} \mu & \begin{array}{ccc}  0&0&0 \end{array} & e 
\\  \begin{array}{c} 0\\0\\0  \end{array}& P_v & \begin{array}{c} 0\\0\\0 \end{array} & P_h
\\ e &  \begin{array}{ccc} 0&0&0 \end{array}  & \gamma & 0
\\  & P_h & 0 & P_h
\end{array}\right)
\]

This will now be our general model for a fluid.
\\

To comply, at last, with more general convention, we note the divergence of a tensor $T$ : $div_g T := \D T$. For a (anti-symmetric) tensor $T=T_{ij}$, the $g$-associated endomorphism is $^eT=T^i\, _j$. Thus now, $^eF=-2\nabla_j Y^i=-2DY$. With computation analogous to those of section 6, one can obtain the following theorem, just applying Bianchi identity, and computing $\D F$. 

Note that in this theorem, the pressure $P$ can be split everywhere into $P=P_v + P_h$, to show physical effects due to the three "classical" dimensions, $P_v$, and those due to the extra "hidden or small" dimensions, $P_h$.

\newpage

\begin{theorem}
If $D$ is a fluid domain as above, the following equalities are satisfied : 
\begin{itemize}
\item Energy Conservation Laws : $$\D (\mu X)=\D (\mu X_0)=g(X_0,\D P)$$ Furthermore : $$\mu^2 X(\frac{e}{\mu })=\mu .g(Y, \D P)-e.g(X_0,\D P)$$ and $$g(Y,\D P)=\D (\, ^eP(Y))$$

\item Electric Charge Conservation Law : 
$$\D (e X)=\D (eX_0)=\D (^eP(Y))=g(Y,\D P)$$

\item Motion Equations : 
\begin{itemize}
\item For the fluid : $$\mu D_X X=-\D P-g(X_0,\D P)X$$
\item For the apparent fluid : $$\mu D_{X_0}X_0 =e.\, ^eF(X_0)-pr_{\mathrm{T}^{\bot}}(\D P)$$
\end{itemize}

\item Maxwell Equations : $dF=0$, and $$\D F=e.X_0+1/2|F|_g .Y-\, ^eP(Y)$$
\end{itemize}

\end{theorem}

The only new point is to prove that a 1-dimensional, timelike, eigenspace of $^eG_{H_x}$ is necessarily unique in $H_x$, to prove seriously that, when $Y$ is a Killing vector field, the vector field $X_0$ and the functions $\mu$ and $e$ are invariant under the flow of $Y$. Once again, this theorem is a purely geometrical fact, based on Bianchi identity. Full proof is given in the last section of this paper.

\subsubsection{Special Fluids}
The definitions we are now going to give are here to obtain the classical physics equations of general relativity. As these equation will be given in 5+m dimensions, it is by projection on the apparent space-time $H_x$ than the comparison will have to be made. As in the previous section, if the double fibration can be turned into a Riemannian double submersion, that is if the metric can go down to the quotients, then these equations will be valid on the "classical" 4-dimensional space-time $(M/W)/S^1$. In particular, if $W=\emptyset$, or $m=0$, we recover the results of section 6 (Kaluza-Klein setting).

\newpage

\begin{definition}Special fluids :

\begin{itemize}
\item A fluid domain $D$ is a \emph{perfect fluid} domain, if $^eP(Y)=0$. Note that $^eP(Y)=\, ^eP_h(Y)$.
\item A fluid domain $D$ is a \emph{perfect isentropic fluid} domain, if it is a perfect fluid domain such that at each point $x \in D$, the pressure tensor $P_x$ is proportional to $g_x-(g_x)_{|\mathrm{T}_x}$, that is $P=p(g+X_0 \ts X_0+Y \ts Y)$ for some smooth function $p:D \rightarrow \mathbb{R}$.
\item A fluid domain $D$ is a \emph{pluperfect fluid} domain, possibly electrically charged, if it is a perfect fluid such that $\D P=0$. For example, dust, for which $P=0$.
\item A pluperfect fluid without electromagnetism, is a pluperfect fluid such that $e=0$ and $F=0$.
\end{itemize}

\end{definition}

Applying the previous theorem for general fluids to these special fluids gives:

\begin{theorem}Concerning the special fluids:

\begin{itemize}
\item For a \emph{Perfect Fluid}, the following equations are valid:
\begin{itemize}
\item Energy Conservation Laws : $$\D (\mu X)=\D (\mu X_0)=g(X_0,\D P)$$ Furthermore : $$\mu^2 X(\frac{e}{\mu })=\mu^2 X_0(\frac{e}{\mu })=-e.g(X_0,\D P)$$ 

\item Electric Charge Conservation Law : $$\D (e X)=\D (eX_0)=0$$

\item Motion Equations : 
\begin{itemize}
\item For the fluid : $$\mu D_X X=-\D P-g(X_0,\D P)X$$
\item For the apparent fluid : $$\mu D_{X_0}X_0 =e.\, ^eF(X_0)-\D P-g(X_0,\D P)X_0$$
\end{itemize}

\item Maxwell Equations : $dF=0$, and $$\D F=e.X_0+1/2|F|_g .Y$$ In particular, $$pr_{H_x}(\D F)=eX_0$$
\\

\end{itemize}
Applying these equations to the more special fluids defined above :

\item For an \emph{Isentropic Perfect Fluid} :
\begin{itemize}
\item Energy Conservation Laws : $$\D \mu X_0+p.\D X_0=0$$ Furthermore : $$\mu^2 X(\frac{e}{\mu })=\mu^2 X_0(\frac{e}{\mu })=e.p.\D X_0$$

\item Electric Charge Conservation Law : $$\D (e X)=\D (eX_0)=0$$

\item Motion Equations : $$(\mu +p).D_{X_0}X_0=e.^eF(X_0)-grad _g(p) - X_0(p).X_0$$

\item Maxwell Equations : $dF=0$, and $$\D F=e.X_0+1/2|F|_g .Y$$  In particular, $$pr_{H_x}(\D F)=eX_0$$

These are exactly the equations of general relativity for isentropic charged fluids.
\end{itemize}

\item For an \emph{Pluperfect Fluid (e.g. electrically charged dust)} :
\begin{itemize}
\item Energy Conservation Laws : $$\D (\mu X)=\D (\mu X_0)=0$$ Furthermore : $$X(\frac{e}{\mu })=X_0(\frac{e}{\mu })=0$$

\item Electric Charge Conservation Law : $$\D (e X)=\D (eX_0)=0$$

\item Motion Equations : 
\begin{itemize}
\item For the fluid : $$D_X X=0$$
\item For the apparent fluid : $$\mu D_{X_0}X_0 =e.\, ^eF(X_0)$$
\end{itemize}
That is, $X$ is a geodesic vector field, even if the fluid has an electrical charge. Of course, if $e=0$, $X_0$ is geodesic.
\item Maxwell Equations : $dF=0$, and $$\D F=e.X_0+1/2|F|_g .Y$$
\end{itemize}

\end{itemize}

\end{theorem}

In the motion equations, $D_{X_0}X_0$ and $^eF(X_0)$ are $g$-orthogonal to $Y$, but not necessarily to $W$. If one wants to guaranty that $D_{X_0}X_0$ and $^eF(X_0)$ belong to $H_x$, one can impose the following requirement :

\emph{The submanifolds $W_x$ are parallel along the geodesic circles $S^1_x$ : $\forall x\in D$, $\forall Z \in T_xW_x$, $\forall x'\in S^1_x$, the parallel transport of $Z$ along $S^1_x$ is tangent to $W_{x'}$.}

It is then quickly verified that under this condition, $\forall x\in D$, $^eF(X_0) \in H_x$.

\subsection{Newtonian and electromagnetic potential in 5+m dimensions.}
\begin{definition}
A domain $D$ of $M$ is a \emph{potential domain} if $\forall x \in D$, $G_{H_x}=0$ and $pr_{H_x}\,^eG(Y)=0$.
\end{definition}
These domain are therefore extensions of fluid domain in which the energy density, the charge density and the apparent pressure $P_v$ are null. Thus, there is no canonically defined vector fields $X$ or $X_0$. Only remains the electromagnetic objects $Y$ and $F=dY^*$. The tensor $G$ is then now equal to the hidden pressure, which therefore satisfies $\D P_h=0$.

The above theorems give:

\begin{theorem}
In a \emph{Potential domain} we have the Second Maxwell equation:
$$\D F= 1/2 |F|_g.Y-\, ^eG(Y)$$
In particular, as $pr_H\, ^eG(Y)=0$, we have : $pr_H \D F=0$.

The first Maxwell equation $dF=0$ is obvious as $F=dY^*$.
\end{theorem}

Potential domains are very important as the knowledge of their geodesics gives the motion curves of "test particles" placed in these potentials, when it is considered that their effect on the geometry of the domain can be neglected. Indeed, if one "introduce" a test particle in a potential domain, it can then be considered as a dust fluid where the energy density $\mu $ is not zero, this "spatial" domain being very limited. If we consider that outside this domain, the geometry of the potential domain is not affected, the flow field $X$ of the fluid, defined when $\mu \neq 0$, is a geodesic field according to theorem 11. The curves of the flow $X$ can thus be considered approximately as the geodesics of the potential domain. Furthermore, the quotient $\frac{e}{\mu }$ can be considered as the quotient of the charge by the mass of the test particle. This is a classical method in general relativity for test particles without electrical charge, as for example in the Schwarzschild solution. The remarkable thing in our setting is that this principle now applies even for test objects with an electrical charge, but in a "space-time" of dimension greater or equal to 5. In this case, the apparent trajectory is determined from the apparent field $X_0$, itself being obtained from the geodesic field $X$. Of course, $X=X_0$ when the charge is zero.

We shall now present examples of such domains, with the big advantage of being given with the exact metric tensor $g$. Precise computations of geodesics will be given, obtained with the help of Mapple software, or, better as it if free, SAGE software.

\subsubsection{Preliminary geometrical setting.}
The circle $S^1(\delta)$ is defined as $S^1(\delta)=\mathbb{R}/2\pi \delta \mathbb{Z}$. The natural surjection $ \Pi : \mathbb{R} \rightarrow 2\pi \delta \mathbb{Z}$ gives a natural origin $P=\Pi (0)$, a natural orientation (that of $\mathbb{R}$ carried by $\Pi$), a natural coordinate $u \in ]0, 2\pi[$ for $\T u =\Pi(u) \in S^1(\delta)-\{P\}$, and a metric $g_{S^1(\delta)}$, the metric of $\mathbb{R}$ quotiented by $\Pi$.

We define the torus $T^n(r_1,...,r_n)=S^1(r_1)\times ...\times S^1(r_n)$, which therefore carries by using the above definition for $S^1(\delta)$, an natural origin, a natural coordinate system, and a natural metric, the product metric : $g_{T^n(r_1,...,r_n)}=g_{S^1(r_1)} \times ... \times g_{S^1(r_n)} $.

We shall call these definitions the "standard setting" on $S^1$ or $T^n$.

 Let $\Theta$ be a open set in $\mathbb{R}^4$ and $$\mathcal{C}=\Theta \times S^1(\delta ) \times  T^{n-5}(r_1,...,r_{n-5})$$
$\mathcal{C}$ will be called a standard cell. The standard coordinate on $\mathcal{C}$ will be denoted by $(t,x,y,z,u,v_1,...,v_{n-5})$, where $(t,x,y,z) \in \Theta$. The standard metric on $\mathcal{C}$ is the product metric $g_0 := g_{\Theta} \times (-g_{S^1(\delta)}) \times g_{T^{n-5}(r_1,...,r_{n-5})}$, where $g_{\Theta}$ is the Minkowski metric on $\Theta \subset \mathbb{R}^4$. In standard coordinates, it is written $$g_0=-dt^2+dx^2+dy^2+dz^2-du^2+dv_1^2+...+dv_{n-5}^2$$ This metric will be called the Minkowski metric of the standard cell $\mathcal{C}$. Its signature is everywhere $(-,+,+,+,-,+,...+)$.

We now consider on a standard cell $\mathcal{C}$, the following two objects : 

-A function $V : \mathcal{C} \rightarrow \mathbb{R}$ where $V$ is a function of the variables $(x,y,z)$.

-A 1-form $\gamma := \phi dt +A_1dx+A_2dy+A_3dz$ on $\mathcal{C}$, where the functions $\phi , A_1,A_2,A_3$ are functions of $(t,x,y,z)$.

We will denote by $F$ the 2-form $F:= d\gamma$ on $\mathcal{C}$.

We shall use the classical terminology : $V$ will be called the Newtonian potential, $\gamma$ the electromagnetic potential ($\phi$ the electric potential, $(A_1,A_2,A_3)$ the magnetic potential ), and $F$ the electromagnetic field 2-form.

We now define the pseudo-Riemannian tensor $g_1$ by 
$$g_1:=g_0-2V. \mathcal{N}_1 \ts \mathcal{N}_1+(\gamma \ts \mathcal{N}_2 +\mathcal{N}_2 \ts \gamma)$$
where $\mathcal{N}_1:=dt+dv_1$, and $\mathcal{N}_2:=du+dv_2$.
In standard coordinates, the matrix of $g_1$ is therefore (considering $n=8$) : 
\[
g_1=\left(\begin{array}{cccccccc}
-1-2V & 0 & 0 & 0 & \phi &-2V& \phi & 0 \\
0 & 1 & 0 & 0 & A_1 & 0 & A_1 & 0 \\
0 & 0 & 1 & 0 & A_2 & 0 & A_2 & 0 \\
0 & 0 & 0 & 1 & A_3 & 0 & A_3 & 0 \\
\phi  & A_1 & A_2 & A_3 & -1 & 0 & 0 & 0 \\
-2V & 0 & 0 & 0 & 0 & -2V+1 & 0 & 0 \\
\phi  &A_1 & A_2 & A_3 & 0 & 0 & 1 &0 \\
0 & 0 & 0 & 0 & 0 & 0 & 0 & 1
\end{array}\right)
\]

We also consider the following two particular cases:

$g_N:=g_0-2V. \mathcal{N}_1 \ts \mathcal{N}_1$, that is $g_1$ where $\gamma=0$,

and

$g_E:=g_0+(\gamma \ts \mathcal{N}_2 +\mathcal{N}_2 \ts \gamma)$, that is $g_1$ where $V=0$.
\\

The following results can then be shown :

\begin{itemize}
\item $det(g_1)=det(g_0)=1$
\item the 1-forms $\mathcal{N}_1$ and $\mathcal{N}_2$ are isotropic both for $g_0$ and $g_1$.
\item $\Delta_{g_0}V=\Delta_{g_1}V$, and will therefore be denoted by $\Delta V$.
\item $Ricc_{g_N}=(\Delta V)\mathcal{N}_1 \ts \mathcal{N}_1$.
\item $Ricc_{g_E}=1/2(H.\mathcal{N}_2 \ts \mathcal{N}_2-(\nabla_{g_0}\cdot F) \ts \mathcal{N}_2+\mathcal{N}_2 \ts (\nabla_{g_0}\cdot F) )$
\item $S_{g_N}=S_{g_E}=0$ 
\item $G_N=2(\Delta V)\mathcal{N}_1 \ts \mathcal{N}_1$
\item $G_E=H.\mathcal{N}_2 \ts \mathcal{N}_2-(\nabla_{g_0}\cdot F) \ts \mathcal{N}_2+\mathcal{N}_2 \ts (\nabla_{g_0}\cdot F)$.
\item For any $(P,Q) \in \Theta \times T^{n-5}$, the circle $\{P\}\times S^1(\delta)\times \{Q\}$ is a timelike geodesic both for $g_0$ and $g_1$.
\end{itemize}
Here $\nabla_{g_0}\cdot F=\partial_i(g_0^{ik}F_{kj})$ and  $H:=|F|_{g_0}=g_0^{ik}g_0^{jl}F_{kl}F_{ij}$.

We then consider the vector field $Y$ tangent to these circles, oriented by the standard orientation, and such that $g_1(Y,Y)=-1$, (in fact $Y=\partial_u$). We have the following results :
\begin{itemize}
\item The electromagnetic potential $\gamma$ is the 1-form $g_E$-associated to $Y$, i.e. $\gamma_i = g_{E_{ij}}Y^j$. 
\item $Y$ is a Killing vector field for $g_E$, and $DY=0$ for $g_N$.
\end{itemize}

These results are not difficult, but obviously requires some tedious computations. A mathematical software like Sage is a big help...

\subsubsection{Newtonian Potential}
We consider in this subsection a triple $(\mathcal{D},g,\mathcal{A})$ where $(\mathcal{D},g)$ is isometric, via some $\varphi :\mathcal{D} \rightarrow \mathcal{C}$, to the standard cell $(\mathcal{C},g_N)$ defined in the previous section, and where $\mathcal{A}$ is the completed observation atlas of $\{(\mathcal{D},\varphi)\}$. From now on, we identify $(\mathcal{D},g)$ with $(\mathcal{C},g_N)$.

Considering $(\mathcal{D},g)$ as a potential domain means that we suppose that $\forall x \in \mathcal{D}$, $G_{Nx|H_x}=0$. From the results of the preliminary setting, this means that $\Delta V=0$, that is, $G_N=0$.

We now study the geodesics of this domain.

In the standard coordinate system $(t,x,y,z,u,v_1,...,v_{n-5})$, the Christoffel symbols of $g_N$ are :
$$\Gamma_{1j}^1=\Gamma_{6j}^1=-\Gamma_{1j}^6=-\Gamma_{6j}^6=\Gamma_{11}^j=\Gamma_{16}^j=\Gamma_{66}^j=\partial_j V$$
where $\partial_j:=\partial/\partial x_j$, and $x_j$ is the $j-th$ coordinate in $(t,x,y,z,u,v_1,...,v_{n-5})$. The other symbols are all zero, except of course the "symmetric" $\Gamma_{ij}^k=\Gamma_{ji}^k$.

Let $\alpha (s)=(t(s),x(s),y(s),z(s),u(s),v_1(s),...,v_{n-5}(s))$ be a geodesic of $(\mathcal{D},g)$, parametrized by $s \in \mathbb{R}$.

Using the above Christoffel symbols, it satisfies the following equations :
$$t''+2(x'.\partial_x V+y'.\partial_yV+z'.\partial_z V)(t'+v_1')=0$$
$$v_1''-2(x'.\partial_x V+y'.\partial_yV+z'.\partial_z V)(t'+v_1')=0$$
$$x''+(t'+v_1')^2.(\partial_x V)=0$$
$$y''+(t'+v_1')^2.(\partial_y V)=0$$
$$z''+(t'+v_1')^2.(\partial_z V)=0$$
$$u''=v_2''=...=v_{n-5}''=0$$
where $t'=t'(s)$, $\partial_x V=(\partial_x V)_{\alpha(s)}$, etc...

The first two equations can be written :
$$t''+2(V(\alpha(s)))'.(t'+v_1')=0$$
$$v_1''-2(V(\alpha(s)))'.(t'+v_1')=0$$
Thus, $t''+v_1''=0$, and $t'+v_1'=k$. Up to a reparametrization, we can suppose $k=1$. ($k=0$ is not interesting for us). Then :
$$t''+2(V(\alpha(s)))'=0$$
$$v_1''-2(V(\alpha(s)))'=0$$
and 
$$t'=-2V(\alpha(s))+c$$
$$v_1'=2V(\alpha(s))+1-c.$$
The three next equations give : $x''=-\partial_x V$, $y''=-\partial_y V$, $z''=-\partial_z V$, that is :
$$(x(s),y(s),z(s))''=-(\nabla V)_{\alpha(s)}$$
which is exactly Poisson equation in classical physics when $V$ is a Newtonian potential ($\Delta V=0$) and when $(x(s),y(s),z(s))$ represents the trajectory of a test particle of mass $m$ in such a potential, but here \emph{considering that $s$ is the time parameter.}

However, if we suppose that $V=o(1)$, and if we consider only the geodesics for which $v_1'(s)=o(1)$, that is those for which the speed component along the circle $S^1(r_1)$ is small compared to 1, (the speed of light), we see from the above equations that $t'(s)=1+o(1)$. In this case, this means that the parameter $s$ is very close to the "time" $t$, and that the Poisson equation is almost classically satisfied.

Remarks concerning this "Newtonian potential":

Let us set, in the standard coordinates, $V=-\frac{m}{r}$, where $r=\sqrt{x^2+y^2+z^2}$ and $m$ is a positive constant. The standard cell is then "space-symmetric" for the usual coordinates $(x,y,z)$. We have $\Delta V=0$, and we just saw that for the metric $g_N=g_0+2\frac{m}{r} \mathcal{N}_1 \ts \mathcal{N}_1$, the geodesics $\alpha (s)=(t(s),x(s),y(s),z(s),u(s),v_1(s),...,v_{n-5}(s))$ satisfies (at least those of interest): 
$$(x(s),y(s),z(s))''=-(\nabla V)_{\alpha (s)}=\frac{m}{r(\alpha (s))^2}. $$
We therefore conclude, as in classical mechanics, that the image of these geodesics projected on classical space $(x,y,z)$ are exacly conics for which $(0,0,0)$ is a focal point, and for which Kepler laws are valid, but when considering the parameter $s$ instead of "time" $t$ of the coordinate system, which can differ greatly from $s$ if $V \neq o(1)$, that is if $r$ is close to $0$.

Because $g_N (\partial_t, \partial_t)=-1+2m/r$, the vector field $\partial_t$ is timelike if $r>2m$, null if $r=2m$, and spacelike if $0<r<2m$.
The critical radius $r=2m$ corresponds to the Schwarschild radius, and therefore suggests to compare this Newtonian potential domain to the classical Schwarschild domain which could be defined here to be a standard cell $(\mathcal{C},g_S)$ :
$$\mathcal{C}=\Theta \times S^1(\delta ) \times  T^{n-5}(r_1,...,r_{n-5})$$
where $\Theta =\mathbb{R}\times \mathbb{R}^{3*}$, and $g_S$ is the product metric $g_S=g_{\Theta} \times g_V$, with $g_V$ the standard metric on $V= S^1(\delta ) \times  T^{n-5}(r_1,...,r_{n-5})$, and $g_{\Theta}$ is the classical Schwarschild metric, written in spherical coordinate $(t,r,\varphi, \phi)$ on $\mathbb{R}\times ]2m, + \infty [ \times S^2 \backsim \mathbb{R}\times \mathbb{R}^{3*}$ :
$$g_{\Theta}(t,r,\varphi, \phi)=(-1+2m/r)dt^2+(1-2m/r)^{-1}dr^2+r^2(d\varphi^2+ sin^2\varphi d\phi^2).$$

The purpose of the product $g_S=g_{\Theta} \times g_V$ is only to carry the classical 4-dimensional Schwarschild metric into our 5+m-dimensional setting. We could also consider the Schwarschild domain extended to $0<r\leqslant 2m$.

Let us compare some properties of the domains $(\mathcal{C},g_N)$ and $(\mathcal{C},g_S)$.
\begin{itemize}
\item Both Ricci curvatures : $Ricc_{g_N}$ and $Ricc_{g_S}$ are zero.
\item For $r>>2m$, the geodesics of $(\mathcal{C},g_N)$ and the timelike geodesics of $(\mathcal{C},g_S)$, constant on $V= S^1(\delta ) \times  T^{n-5}(r_1,...,r_{n-5})$, give, with very good approximation, the trajectories of test particles around a body of mass $m$ with spherical symmetry in $(x,y,z)$ space, computed in classical Newtonian mechanics.
\end{itemize}

One can also notice that the coefficient $(-1+2m/r)$ in front of $dt^2$ in $g_N$ is the same as that of $g_S$. However, for $g_N$ the potential $2m/r$ is perturbating the "small" dimensions of $T^{n-5}$ without affecting the classical dimensions $(x,y,z)$, whereas for $g_S$, the potential $2m/r$ perturbates the $(x,y,z)$ dimensions, without affecting the "small" dimensions.

The two cells $(\mathcal{C},g_N)$ and $(\mathcal{C},g_S)$ could therefore be considered as extreme particular cases of a family of domains resembling Newtonian potentials, for which the potential is affecting every dimensions, and which still satisfy the two above conditions.

\subsubsection{Electromagnetic Potential}
We now consider a triple $(\mathcal{D},g,\mathcal{A})$ where $(\mathcal{D},g)$ is isometric, via some $\varphi :\mathcal{D} \rightarrow \mathcal{C}$, to the standard cell $(\mathcal{C},g_E)$ defined in section 7.4.1, and where $\mathcal{A}$ is the completed observation atlas of $\{(\mathcal{D},\varphi)\}$. From now on, we identify $(\mathcal{D},g)$ with $(\mathcal{C},g_E)$.

By the given definition of a potential domain, and the results of section 7.4.1, $\nabla_{g_0} \cdot F=0$. Thus : 
$$G_E=H. \mathcal{N}_2 \ts \mathcal{N}_2.$$

 Let us study the geodesics of this domain.

In the standard coordinate system $(t,x,y,z,u,v_1,...,v_{n-5})$, to simplify the numeration of the Christoffel symbols, we set $\mathcal{N}_2=du+dv_1$, (instead of $du+dv_2$). In this coordinate system, the Christoffel symbols of $g_E$ satisfy :
\begin{itemize}
\item $\forall k$, $\Gamma_{ij}^k=0$ if $i,j \geqslant 5$, $i$ or $j$ $>6$.
\item $\forall i,j$, $\Gamma_{ij}^k=0$ if $k>6$.
\item $\Gamma_{ij}^k=0$ if $i,j,k<5$.
\item  $\Gamma_{i5}^k=\Gamma_{i6}^k=1/2g^{kk}(\partial_i g_{5k}-\partial_k g_{i5})=1/2g^{kk}(\partial_i g_{6k}-\partial_k g_{i6})$ if $k<5$.
\item $\forall i,j$, $\Gamma_{ij}^5+\Gamma_{ij}^6=0$.
\end{itemize}

Let $\alpha (s)=(t(s),x(s),y(s),z(s),u(s),v_1(s),...,v_{n-5}(s))$ be a geodesic of $(\mathcal{D},g)$, parametrized by $s \in \mathbb{R}$.
Using the above Christoffel symbols, it satisfies the following equations :

If $k<5$, (i) : $x_k'' (s)+2(\sum_{i=1}^4 \Gamma_{i5}^k x_i' (s))(x_5'(s)+x_6' (s))=0$, where $x_j$ is the $j-th$ coordinate in $(t,x,y,z,u,v_1,...,v_{n-5})$, and $\Gamma_{i5}^k=\Gamma_{i5}^k (\alpha(s))$.

Furthermore, $u''+\sum_{i,j}\Gamma_{ij}^5 x_i'x_j'=0$, and $v_1''+\sum_{i,j}\Gamma_{ij}^6 x_i'x_j'=0$.

From this we get : $(u+v_1)''(s)=0$, and $u'+v_1'=c$. (i) can therefore be rewritten : $x_k''=-2c(\sum_{i=1}^4 \Gamma_{i5}^k x_i' (s))$.

But $Y$ is a Killing vector field, so $F^k \, _i=-2\nabla_i Y^k=-2(\partial_i Y^k +\Gamma_{il}^k Y^l)=-2\Gamma_{i5}^k$. So : 
$$x_k''(s)= c \sum_{i=1}^4 F^k \, _i x_i'(s)$$
as $F^k \, _i=0$ if $i\geqslant 5$.

Now, $\alpha(s)$ can be parametrized so that $g(\dot \alpha(s),\dot \alpha(s))=-1$, and $\dot \alpha(s)_{|H_x}$ is in the time orientation given by $t$. We will of course call $s$ the proper time of the geodesic.

Denoting by $X(s)=(t'(s),x'(s),y'(s),z'(s))$ the vector corresponding to the first 4 components of $\dot \alpha(s)$, the above equations can be written : 
$$X'(s)=c.\, ^eF(X(s)).$$

We recover the classical equation of the motion of a particle of mass $m$ and of charge $q$ in an electromagnetic field $F$ when we set $c=q/m$ and when $s$ indeed represents the proper time of the particle.

Here, $c=q/m=u'+v_1'$ is a caracteristic data of the geodesic on the "small" dimensions.

If the speeds $|x_k '(s)|=o(1)$ for $k \neq 1,5,6$ and if $|u'|=|v_1'|+o(1)$, then $t'=1+o(1)$ and the "time" $t$ of the coordinate system is very close to the proper time of the geodesics; this corresponds to the classical approximation on non relativistic classical mechanics.

\subsection{Example of perfect fluid without electromagnetism, static, without visible pressure, but with hidden pressure.}
Here the fluid domain considered is a triple $(\mathcal{D},g,\mathcal{A})$ where $(\mathcal{D},g)$ is isometric, via some $\varphi :\mathcal{D} \rightarrow \mathcal{C}$, to $(\mathcal{C},g)$ where this latest is defined in the following way :

$\mathcal{C}=\Theta \times S^1(\delta ) \times  T^{n-5}(r_1,...,r_{n-5})$
is a standard cell equipped with standard coordinate $(t,x,y,z,u,v_1,...,v_{n-5})$. 

$g=g_0+\beta \ts \mathcal{N}_1 +\mathcal{N}_1 \ts \beta$, where $\mathcal{N}_1=dt+dv_1$, and $\beta =a.dx+b.dy+c.dz$.
$a,b,c: \mathcal{C} \rightarrow \mathbb{R}$ are smooth functions depending only on the variables $x,y,z$.

Note that the metric does not depend on $t$, which is why it can be called "static".

Notice that although the metric $g$ seems alike the metric $g_E$ of the electromagnetic potential, we use $\mathcal{N}_1=dt+dv_1$ instead of $\mathcal{N}_2:=du+dv_2$, which leads to a fundamental difference. We can use the computation made for $g_E$ though, the role of $u$ and $t$ being exchanged. More simply, a mathematical software gives the following results.

We denote by $(A,B,C):=curl(a,b,c)$, that is :
$$A=\partial b/\partial z - \partial c/\partial y$$
$$B=\partial c/\partial x - \partial a/\partial z$$
$$C=\partial a/\partial y - \partial b/\partial x$$
We note $(\mathcal A, \mathcal B, \mathcal C):=curl(A,B,C)=curl.curl(a,b,c) $.
We obtain the following:
\[
Ricc_g=\frac{1}{2}\left(\begin{array}{ccccccc}
(A^2+B^2+C^2) & \mathcal A & \mathcal B & \mathcal C & (A^2+B^2+C^2)& 0 &0 \\
\mathcal A & 0 & 0 & 0 & \mathcal A & 0 &0 \\
\mathcal B & 0 & 0 & 0 & \mathcal B & 0 &0  \\
\mathcal C & 0 & 0 & 0 & \mathcal C & 0  &0 \\
(A^2+B^2+C^2) & \mathcal A & \mathcal B & \mathcal C & (A^2+B^2+C^2) & 0 &0 \\
0 & 0 & 0 & 0 & 0 & 0 &0 \\
0 & 0 & 0 & 0 & 0 & 0 &0 
\end{array}\right)
\]
(This is written for $n=7$ ; for $n$ greater, the rest of the matrix is filled with zeros.) Also, 
$$S_g=0$$
To obtain a pluperfect fluid, we impose the two following conditions :
$$(\mathcal A, \mathcal B, \mathcal C)=curl.curl(a,b,c)=0$$ and $$\mu :=A^2+B^2+C^2\neq 0$$
Then $$G_g=\mu \, \mathcal N_1 \ts \mathcal N_1$$
where $\mu =||curl(a,b,c)||^2=A^2+B^2+C^2>0$.

In the standard coordinate system, the matrix of $g$ is :
\[
g_{ij}=\left(\begin{array}{ccccccc}
-1 & a & b & c & 0& 0 & 0 \\
a & 1 & 0 & 0 & 0 & a & 0\\
b & 0 & 1 & 0 & 0 & b & 0  \\
c & 0 & 0 & 1 & 0 & c & 0  \\
0 & 0 & 0 & 0 & -1 & 0 & 0 \\
0 & a & b & c & 0 & 1 & 0 \\
0 & 0 & 0 & 0 & 0 & 0 & 1
\end{array}\right)
\]

For every $x \in \mathcal C$, the apparent space-time $H_x$ has the following $g$-orthonormal basis :
$$ X_0=(1,0,...,0)$$
$$ X_1=(a,1,0,0,0,-a,0,...,0)$$
$$X_2=(b,0,1,0,0,-b,0,..,0)$$
$$X_3=(c,0,0,1,0,-c,0,...,0)$$
As $G(X_0,X_0)=\mu$ and $G(X_i,X_j)=0$ if $(i,j) \neq (0,0)$, the matrix of $G_{|H_x}$ in this basis is :
\[
\left(\begin{array}{cccc}
\mu  & 0 & 0 & 0  \\
0 & 0 & 0 & 0 \\
0 & 0 & 0 & 0   \\
0 & 0 & 0 & 0 
\end{array}\right)
\]
From the given definitions, the vector field $X_0=\partial_t$ is the apparent field of the fluid, and $\mu$ is the energy density function, (here $X_0=X$ is the field of the fluid).

The visible pressure is zero. The hidden pressure $P_h$ is then the pressure $P$, and $P=G-\mu X_0^\flat \ts X_0^\flat $. One can check that $\D P=0$ ; it suffices, as $\D G =0$, to check that $\D (\mu X_0^\flat \ts X_0^\flat)=0$. 

To do that, let us set $ X_0 \ts X_0 = T^{ij} \partial_i \ts \partial_j$, with $T^{11}=1$, and $T^{ij}=0$ if $(i,j) \neq (1,1)$. One has :
$$\nabla_i \mu T^{ij}=\partial_i (\mu T^{ij})+\mu (T^{lj}\Gamma^i_{il}+   T^{il}\Gamma^j_{il})=\mu (T^{1j}\Gamma^i_{i1}+\Gamma^j_{11})$$
as $\partial_i (\mu T^{ij})=\partial_1 (\mu T^{11})=\partial_1 \mu =0$, ($\mu$ does not depend on $t$).

Besides, $\Gamma^j_{11}=-g^{jl}(\partial_l g_{11})=0$ because $g_{11}=-1$. Then:

$\Gamma^i_{i1}=1/2g^{il}(\partial_ig_{l1}+\partial_1g_{li}-\partial_lg_{i1})=1/2g^{il}(\partial_ig_{l1}-\partial_lg_{i1})=0$ because $g^{il}$ is symmetric in $i,l$ and $(\partial_ig_{l1}-\partial_lg_{i1})$ is anti-symmetric.

The domain $(\mathcal{D},g,\mathcal{A})$ is therefore a \emph{pluperfect fluid domain}, the other conditions being quickly checked. Its visible pressure, is zero, and the vector field of the fluid, $X_0=X=\partial_t$, is a geodesic vector field, for $g$ and $g_0$.
\\

\emph{Remark} : If one consider as in example 7.4.2., a Newtonian potential, without the requirement $\Delta V=0$, one can check that it gives a fluid type domain, but that this fluid is not "perfect".

\subsection{An geometric approach to quantum mechanics in 5+m dimensions by Michel Vaugon.}

Building on the geometric setting described in this paper, Michel Vaugon gives in the following manuscript "relavite quantique michel.juillet2012", available on the dropbox page :
\\

https://www.dropbox.com/sh/s0wdvk90mwxneuf/yOh5cK76Zc
\\

a possible way to introduce classical quantum mechanics in our setting, by describing types of domain called "metric waves", which give with precise approximations the classical Schrodinger and Klein-Gordon equations.

These manuscripts are under translation and Latex typing... They will be available as an extension of this paper in the near future...

\newpage

\section{Proof of theorem 9.}
It is essentially the same computations as in section 6, but in the more general setting of dimension $5+m$, and with $Y$ now being timelike. Once again, this theorem is a purely geometrical fact. We keep notation of section 7.3.1.
\\

The first point is to prove that a 1-dimensional, timelike, eigenspace of $^eG_{H_x}$ is necessarily unique in $H_x$, which is very elementary. However this is necessary to prove seriously that, when $Y$ is a Killing vector field, the vector field $X_0$ and the functions $\mu$ and $e$ are invariant under the flow of $Y$.
\\

Suppose that $X_0$ and $X_0'$ are both timelike eigenvectors of $^eG_{|H_x}$ of respective eigenvalue $\lambda$ and $\lambda'$. Then $G(X_0,X_0')=\lambda g(X_0,X_0')$ on the one hand, and $G(X_0',X_0)=\lambda' g(X_0',X_0)$ on the other hand. But, because of the signature of $g_{|H_x}$, $g(X_0,X_0') \neq 0$, therefore $\lambda =\lambda'$, and thus $X_0'$ is a eigenvector for $\lambda$, proportional to $X_0$.
\\

Fix now a point $P$ in $D$, and consider the fibers $S^1_P$ and $W_P$ as defined in section 7.2.2. Fix a $(S_1 \times W)$-chart $\phi : \mathcal{V}(P) \rightarrow \Theta \times S^1 \times W$ on a neighborhood $\mathcal{V}(P)$ of $P$, and pick coordinates charts $(x^0,x^1,x^2,x^3)$ on $\Theta$, $(u)$ on $S^1$ and $(w^1,...,w^m)$ on $W$ so that $(x^0,...,x^3,u,w^1,...,w^m)$ gives coordinates around $\phi (P)$. If $\gamma :I \rightarrow S^1_P$ is a path whose image is contained in the fiber $S^1_P$ at $P$, its coordinates in the above chart can be written  $(x^0(t),...,x^3(t),u(t),w^1(t),...,w^m(t))$. But, as by definition, $$S_P^1=\phi^{-1}(\phi^{\Theta}(P) \times S^1 \times \phi^W(P)),$$ necessarily $x^i(t)=cst=x^i(P)$ for $i=0,...,3$, and $w^i(t)=cst=w^i(P)$ for $i=1,...,m$.

Let us denote by $\sigma_s$ the 1-parameter group of diffeomorphisms generated by the flow of $Y$ (note that this flow is complete as $Y$ is Killing). Considering, for any $x$ in $D$, the path $s \mapsto \sigma_s(x)$, we see from the above remark concerning a path whose image is included in $S^1_p$, and using coordinates as defined above around $P$, that, for $s$ small enough, and for $x$ in a small neighborhood of $P$, $\sigma_s$ can be written:
$$\sigma_s (x^0,...,x^3,u,w^1,...,w^m)=(x^0,...,x^3,\sigma_s^5(x^i,u,w^i),w^1,...,w^m)$$
The Jacobian matrix of the tangent map $\sigma_{s*,P}$ of $\sigma_s$ at $P$ will therefore be, in these coordinates, of the form:
\[
Jac(\sigma_{s*,P})=\left(\begin{array}{ccc} Id_{\mathbb{R}^4} &\begin{array}{c}0\\ \vdots \\0 \end{array}& 0 \\ 
\partial_{x^0}\sigma_s^5 \cdots \partial_{x^3}\sigma_s^5 &\partial_{u}\sigma_s^5 & \partial_{w^1}\sigma_s^5 \cdots \partial_{w^m}\sigma_s^5 \\
0&\begin{array}{c}0\\ \vdots \\0 \end{array}& Id_{\mathbb{R}^m} 
\end{array}\right)
\]
where $Id_{\mathbb{R}^4} $ is the identity matrix. This shows that 
$$\sigma_{s*,P}(T_P(S^1_P)\times T_PW) \subset T_{\sigma_s(P)}S^1_P \times T_{\sigma_s(P)}W_{\sigma_s(P)}.$$ 
(Of course, by definition, $S^1_P=S^1_{\sigma_s(P)}$). In particular, for any $Z \in T_PW_P$, $\sigma_{s*,P}(Z) \in T_{\sigma_s(P)}S^1_P \times T_{\sigma_s(P)}W_{\sigma_s(P)}$. Besides, by definition, $\sigma_{s*}(Y)=Y$. By continuity and additivity in $s$, this can be extended to any value of $s$.

Because the $\sigma_s$ are isometries, and because at each $P \in D$ $H_P$ is defined as $H_P=(T_P(S^1_P)\times T_PW)^{\bot}$, we conclude that for any $s$ :
$$\sigma_{s*,P}(H_P)=H_{\sigma_s(P)}$$
Therefore, by the unicity of the 1-dimensional timelike eigenspace of $^eG_{H_x}$ proven at the beginning, we get that $\forall s \in \mathbb{R}$, $\forall x \in D $, 
$$\sigma_{s*,P}(X_0(P))=\pm X_0(\sigma_s(P))$$
In an $(S^1 \times W)$-chart, by the chosen orientation, for any point $P$, $g(X_0(P), \phi^*(\partial_{x^0}))<0$. So for any $s$, 
$$g(X_0((\sigma_s(P)), \phi^*(\partial_{x^0}))=\pm g(\sigma_{s*,P}(X_0(P)), \phi^*(\partial_{x^0}))<0$$
But for $s=0$, $\sigma_0=Id$, so the $\pm$ sign above must be $+$. We conclude that $\forall s \in \mathbb{R}$, $\forall x \in D $, 
$$\sigma_{s*,P}(X_0(P))= X_0(\sigma_s(P)).$$
$X_0$ is therefore invariant by the isometries $\sigma_s$, as are $Y$ and the Einstein curvature $G$. By their geometric definitions from these, the functions $\mu $, $e$ and $\beta$ are also : $Y(\mu )=Y(e)=Y(\beta)=0$.
\\

Then we prove basic properties of the vector field $Y$: Just suppose that $Y$ is defined as the natural unit tangent vector field to the fiber $S_x$ at each point, with the natural orientation induced by the $S^1$-atlas, with the hypothesis that $Y$ is a Killing vector field. Then:

We just proved that $X_0$ is invariant by the flow of $Y$, This means that the Lie derivative $L_Y X_0=D_Y X_0-D_{X_0}Y=0$. Thus :
$$ D_Y X_0= D_{X_0} Y$$ 
Now, $Y$ being Killing, $\nabla_i Y_j=-\nabla_j Y_i$. So in particular, $\nabla_iY^i=\D Y=0$ :
$$\D Y =0$$ 
As $g(Y,Y)=Y^iY_i=-1$, we have $0=\nabla_j(Y^iY_i)=2Y^i\nabla_j Y_i=-2Y^i\nabla_iY_j$. So :
$$D_Y Y=0$$ 
At last, $F_{ij}=\nabla_iY_j-\nabla_jY_i=2\nabla_iY_j=-2\nabla_jY_i$. That is :
$$^eF=-2DY$$

Let us recall that all is based on the following tensor field :
\begin{equation*}
\fbox{ $\begin{split}
G &=\mu X \ts X+ \beta Y\ts Y+P\\
       & =\mu X_0 \ts X_0+ e(X_0 \ts Y+Y \ts X_0)+(\beta+\frac{e^2}{\mu}). Y \ts  Y+P
       \end{split}$}
\end{equation*}

The first (easy) step is to compute $\D G$ which is zero by Bianchi identity.

Using the fact that $Y$ is geodesic and Killing, which implies, as we saw above, that $\D Y=0$ and $D_Y Y=0$, and $Y(\beta)=Y(e)=Y(\mu)=0$, one obtain:

$$0=\D G= X(\mu ).X+\mu.( \D X).X+ \mu D_X X + \D P$$

concerning the first expression of $G$. And concerning the one using $X_0$,
\begin{equation*}
 \begin{split}
0=\D G &=X_0(\mu ).X_0+\mu D_{X_0}X_0+ \mu (\D X_0).X_0\\
       & +X_0(e).Y+e(\D X_0).Y-e.\,^eF(X_0)+\D P\\
       &=\mu D_{X_0}X_0+(\D(\mu  X_0)).X_0+(\D (eX_0)).Y-e.\,^eF(X_0)+\D P
       \end{split}
\end{equation*}
where we used $D_{X_0}Y=D_Y X_0=-1/2.^eF(X_0)$.

We note $(V,W)$ the scalar product $g(V,W)$ of two vector $V$ and $W$. Using the fact that $(X_0,X_0)$, $(Y,Y)$, $(X_0,Y)$ are constant and that $D_Y Y=0$, it is easy to get that $(D_{X_0}X_0,Y)=(D_Y X_0,X_0)=(D_Y X_0,Y)=0$ and thus that 
 $(^eF(X_0),X_0)=(^eF(X_0),Y)=0$. (Review section 6 for detailed computations).
 
 We now compute the scalar product of $\D G$, seen as a vector, with $X_0$ and $Y$; this will give the conservation laws. The projection of $\D G$ on $H$ will give the motion equation.

$(\D G,X_0)=-\D (\mu X_0)+(\D P, X_0)$

Now $X=X_0+\frac{e}{\mu }Y$, therefore $(X,X)=-(1+\frac{e^2}{\mu^2 })=cst$. Thus as above, $(D_X X, X)=0$. Besides, $\D (\mu X)= \D (\mu X_0)+\D (eY)$, but $\D (eY)=Y(e)Y+eD_YY=0$. This lead to:
$$\D (\mu X)= \D (\mu X_0)=+(\D P, X_0)$$

In the same way :
\begin{equation*}
 \begin{split}
(\D G,Y) &=-X_0(e ).X_0-e. \D X_0+(\D P, Y)\\
       &= -\D (eX_0)+(\D P,Y)\\
       &=0
       \end{split}
\end{equation*}
But just as above, $\D (eX)=\D (eX_0)+\D (\frac{e^2}{\mu^2 }Y)$, so
$$\D (eX)=\D (eX_0)=(\D P, Y).$$

We now want to obtain another expression for $(\D P, Y)$. Using coordinates, we have that:
$$(\D P, Y)=Y^i \nabla_j P_i\,^j=Y^i\nabla_j P^j\, _i$$
as $P$ is symetric. On the other hand, $^e P(Y)=P^i\,_j\, Y^j$. So :
\begin{equation*}
 \begin{split}
\D (^e P(Y))=\nabla _i (P^i\,_j\, Y^j)&=Y^j \nabla _i P^i\,_j +P^i\,_j\,\nabla_iY^j \\
       &=(Y,\D P)+P^i\,_j\,\nabla_iY^j \\
       \end{split}
\end{equation*}
Now, $\nabla_i Y^j=-\nabla_j Y^i $ because $Y$ is Killing, (i.e. $DY$ is anti-symmetric), and on the other hand, $P^i\, _j=P^j\, _i$ as $P$ is symmetric. Therefore, 
$$P^i\, _j\nabla_i Y^j= - P^j\, _i \nabla_j Y^i,$$
thus, $P^i\, _j\nabla_i Y^j=0$; and so : 
$$(Y,\D P)= \D (^eP(Y)).$$

Concerning the conservation law for $\frac{e}{\mu}$, we have 
$$\D (\mu X)=X(\mu )+ \mu \D X=(\D P,X_0)$$ and
$$\D (e X)=X(e )+e \D X=(\D P,Y)$$
Multiplying the first equation by $e$, and the second by $\mu $, then subtracting, one get :
$$\mu^2 X(\frac{e}{\mu})=\mu (\D P,Y)-e(\D P,X_0).$$

Now we pass to the motion equations. These are just the Bianchi identity. First :
$\D G=0=\D (\mu X).X+ \mu D_X X+ \D P$, and $\D (\mu X)=(\D P, X_0)$. Therefore : 
$$\mu D_X X= -\D P- (\D P, X_0).X.$$

For the apparent fluid, we saw that :
$$0=\D G=\mu D_{X_0}X_0 +(\D (\mu X_0)).X_0+(\D (eX_0)).Y-e.\, ^eF(X_0) + \D P$$ so
$$\mu D_{X_0}X_0 +(X_0,\D P).X_0+(Y, \D P).Y-e.\, ^eF(X_0) + \D P =0$$
that is :
$$\mu D_{X_0}X_0=e.\, ^eF(X_0)-pr_{\mathrm{T}^{\bot}}(\D P).$$
\\

At last, we compute $\D F$, using coordinates :

By definition $F=d(Y^*)=\nabla_i Y_j-\nabla_j Y_i$. Then $^eF=\nabla^i Y_j-\nabla_j Y^i=F^i\, _j$.
But $Y$ is Killing, so that $\nabla^i Y_j + \nabla_j Y^i=0$. Thus : $^eF=-2\nabla_j Y^i=F^i\, _j$.

Now, $\D F=\nabla_i F^{ij}=-2\nabla_i \nabla^j Y^i$. But from (one of) the Bianchi identities:
$$\nabla_i \nabla_j Y^k -\nabla_j \nabla_i Y^k=R^k\, _{lij}Y^l$$
we obtain:
$$\nabla_i \nabla^j Y^i-\nabla^j \nabla_i Y^i=R^j\, _l Y^l$$
so $$\D F=-2\nabla^j \nabla_i Y^i-2R^j\, _l Y^l=-2R^j\, _l Y^l$$
as $\nabla_i Y^i=\D Y=0$.

But, $2R^j\, _l =G^j\, _l +S_g.\delta^j\, _l \,$, so $2R^j\, _l Y^l=G^j\, _l.Y^l +S_g.Y^j$. 

Here, $G=\mu X_0 \ts X_0+ e(X_0 \ts Y+Y \ts X_0)+\gamma. Y \ts  Y+P$.

So, $G(Y)=-eX_0-\gamma  Y+P(Y)$. We therefore obtain :
$$\D F= eX_0+(\gamma -S_g).Y-P(Y).$$
Now, this gives $g(Y,\D F)=-\gamma+S_g$, as $g(P(Y),Y)=P(Y,Y)=0$. On the other hand, $0=D_Y Y=Y^j\nabla_j Y^i$, which leads to : 
\begin{equation*}
 \begin{split}
0=\nabla_i(Y^j\nabla_jY^i) &=(\nabla_i Y^j)(\nabla_j Y^i)+Y^j(\nabla_i \nabla_j Y^i) \\
       &=-1/4(F^j\, _i )(F^i\, _j )+Y^j (\nabla_i \nabla_j Y^i)  \\
       &=-1/4(F^j\, _i )(F^i\, _j )-1/2g(Y,\D F)
       \end{split}
\end{equation*}
that is : $g(Y,\D F)=-1/2 |F|_g$. So : 
$$\D F=e.X_0+1/2|F|_g.Y-P(Y)$$
$$\gamma = 1/2|F|_g+S_g$$

\newpage

\section{Help and notations.}
We put here some of our notations and conventions.

In all the paper $M$ (or $\B M$) is a Lorentzian manifold equipped with a metric $g$ (or $\B g$). For $x \in M$, we denote by $T_x M$ the tangent space to $M$ at $x$.

A subset $\Omega \subset M$ is called a domain if it is a open connected subset of $M$.

We will always note $D$ the covariant derivative associated to the  Levi-Civita connexion of $g$. Therefore, for vector fields $X,Y$ on $M$, we note $D_X Y$ the covariant derivative of $Y$ along $X$, and for a tensor $T$, we note $DT$ the covariant differential. We therefore keep the notation $\nabla$ to the indices writing. We in fact use the index notation "a la Wald", \cite{W}, meaning that $T^{ij}$, for example, does not refer to the use of a particular chart, but is used to indicate the type of the tensor $T$, and the possible contractions. Then $\nabla _i T^{kl}$ is, according to this convention, the type of the tensor $DT$ if $T=T^{kl}$, or, if a chart is given, the component $U_i\,^{kl}$ of the tensor $U=DT$.

If there is no ambiguity, we identify, for a rank 2 tensor, $T^{ij}$ and $T_{ij}$. For such a tensor, we note, in section 6, $^eT$ the endomorphism associated by $g$, $^eT=T_i \,^j$. But, for historic reasons, in section 7, it will be $^eT=T^i\,_j$. Caution, if $T$ is not symmetric, one has to say which index is raised (or lowered).

If $f$ is a function and if  $X,Y$ are vector fields on $M$, for a contraction $c$ on the first indices, we have in intrinsic writing, $$div_g(f.X\otimes Y):=c(D(f.X\otimes Y))=D_X f. Y+ f.div_gX.Y+f.D_X Y.$$
where $div_g:=\D$ is the divergence operator.

We note $R_{ij}$ the Ricci curvature, and $R=R^i_i$, or $S_g=R^i_i$ the scalar curvature of $g$.

For vectors $X,Y$, we note $<X,Y>$ the subspace generated by $X$ and $Y$. For a subspace $E$ of the tangent space $T_xM$, we note $E^{\bot}$ the subspace $g_x$-orthogonal to $E$.

At last, concerning fluids of matter, following classical usage, for example \cite{HE} or \cite{W}, a perfect fluid in the classical setting of general relativity is a domain of the space-time manifold where the energy-momentum tensor can be written $T^{ij}=\mu X^iX^j+\lambda(X^iX^j+g^{ij})$, or $T= \mu X \ts X+ \lambda(X \ts X+Id)$. Then, for us, a v-perfect fluid is a perfect fluid for which $\lambda=0$, i.e. $T^{ij}=\mu X^iX^j$, which is usually called a "dust" fluid.
\newpage

\enddocument